\documentstyle[preprint,prb,aps]{revtex}
\begin{document}
\input{psfig}
\draft
\title{Lifting of Multiphase Degeneracy by Quantum Fluctuations} 
\author{
A. B. Harris,$^{1,2}$ C. Micheletti$^1$ and J. M. Yeomans$^1$ }
\address{ (1) Theoretical Physics, Oxford University,
1 Keble Rd. Oxford OX1 3NP, UK}
\address{(2) School of Physics and Astronomy, Raymond and Beverly
Sackler Faculty of Exact Sciences,\\
Tel Aviv University, Tel Aviv 69978, Israel}
\date{\today}
\maketitle
\begin{abstract}
We study the effect of quantum fluctuations on the multiphase point of
the Heisenberg model with first- and second-neighbor competing
interactions and strong uniaxial spin anisotropy $D$. By studying the
structure of perturbation theory we show that the multiphase
degeneracy which exists for $S=\infty$ (i.e., for the ANNNI model)
is lifted and that the effect of quantum fluctuations is to
stabilize a sequence of phases of wavelength 4,6,8,...~.
This sequence is probably an infinite one.  We also show that quantum
fluctuations can mediate an infinite sequence of layering transitions through
which an interface can unbind from a wall.
\end{abstract}

\pacs{PACS numbers: 75.30.Et, 71.70.Ej, 75.30.Gw}

\section{Introduction}

There are many naturally occurring examples of uniaxially modulated
structures. The ferrimagnetic states of the rare
earths [\onlinecite{JJ}] include several 
phases where the wavevector lies along the $\hat{\rm c}$-axis and can be of a
long period commensurate or incommensurate with the underlying
lattice.  Modulated atomic ordering has been observed in metallic
alloys such as TiAl$_3$ and a relationship established between the
wavelength of the modulated phases and the temperature[\onlinecite{alloys}]. Polytypism
describes the phenomenon whereby a compound can have modulated
structural order of different periods [\onlinecite{politypism}]. A well-known example is SiC
where the `ABC' stacking sequence of the close-packed layers can
correspond to many varied and often very long wavelengths.
These systems have been usefully modeled in terms of arrays of
interacting domain walls [\onlinecite{MEFXS}]. When the wall energy is
small, wall--wall 
interactions become important in determining the wall spacing and
small changes in the external parameters can lead to many different
modulated phases becoming stable. 

A model which has proved very useful
for understanding this process is the axial next-nearest neighbor
Ising or ANNNI model which is an Ising system with first- and
second-neighbor  competing interactions along one lattice direction
[\onlinecite{RJE}]. At zero temperature the ANNNI model has a
multiphase point where an infinite number of phases are
degenerate corresponding to zero domain wall
energy. At
low temperature entropic fluctuations cause domain wall interactions
which stabilize a sequence of modulated structures [\onlinecite{BAK}-\onlinecite{VG}]. Our aim in this paper
is to investigate whether quantum fluctuations can play a similar
r\^{o}le. That quantum fluctuations can remove ground state
degeneracies not required by symmetry was pointed out by
Shender [\onlinecite{EFS}] and
termed ``ground state selection'' by Henley [\onlinecite{CLH}].

We find that quantum fluctuations do indeed remove the infinite
degeneracy of the multiphase point of the ANNNI model. A sequence of
first order transitions is stabilized in a way qualitatively similar to
the finite temperature behavior but involving a different sequence
of phases. However, for long-period phases entropic and quantum
fluctuations behave in a subtly different way.

Our analysis focuses on the domain wall interactions and we calculate
in turn the wall energy, two-wall interactions and three-wall
interactions [\onlinecite{MEFXS}]. This is done by an analysis of the
structure of perturbation theory around the multiphase point of the
ANNNI model: all orders of perturbation theory are important. The
calculation is described in Secs. 3 and 4 and corrections pertinent to the
long-period phases are treated in Sec. 5.

To illustrate the essence of the phenomenon,
we start, in Sec. 2, by focussing on a simpler problem 
concerning the unbinding of a single interface.  In this
model, which is effectively a one-wall version of the ANNNI problem,
spins at opposite sides of the system are fixed to be antiparallel.
When the magnetic field, $h$, is nonzero, the domain wall separating
up spins from down spins is bound to one of the surfaces.
For $h=0$ and for an Ising model, there is a
multiphase degeneracy, because the interface energy is independent
of its distance from the surface.  However, when the Ising model is replaced 
by a very
anisotropic Heisenberg model, then, as we show, quantum fluctuations 
induce a surface--interface repulsion
resulting in the interface's unbinding through a series of first order
layering transitions [\onlinecite{GO}]. This calculation is similar in
spirit, but much simpler than that considered in the rest of the paper.

\section{Interface Unbinding Transition}

Our first aim is to show how quantum fluctuations can affect the
unbinding transition of an interface from a surface. Accordingly, we
consider the 
Hamiltonian 
\begin{equation}
{\cal H} =
 - {J \over S^2} \sum_{i=1}^{N-1} {\bf S}_{i} \cdot {\bf S}_{i+1}
+ {h \over S} \sum_{i=2}^{N-1}S_{i}^z 
  -  {D \over S^2} \sum_{i=1}^N ([S_{i}^z]^2 -S^2) 
-{H \over S} (S_1^z-S_N^z) ,
\label{HAMIL}
\end{equation}
where $i$ are the sites of a one-dimensional lattice of length $N$ and
${\bf S}_i$ is a quantum spin of magnitude $S$ at site $i$.
In Eq. (\ref{HAMIL}) we introduced factors of $S$
to simplify the classical spin ($S \rightarrow \infty$) limit.
Although the results are described for one dimension, they hold
for any dimension because of the translational invariance of the
interface parallel to the surface (walls are flat in two or more
dimensions for an Ising model at sufficiently low temperature).
The final term is chosen to impose the boundary conditions such that
there is an interface in the system. The interface will be defined as
being in position $k$ when it lies between sites $k$ and $k+1$.
We shall restrict ourselves to the 
limits of zero temperature, $H=\infty$ and $N=\infty$. 

For $D=\infty$, $S_i^z=\sigma_i S$ where $\sigma_i=\pm 1$ and the
Hamiltonian~(\ref{HAMIL}) reduces to an Ising model in a magnetic
field
whose Hamiltonian ${\cal H}_I$ is
given by
\begin{equation}
\label{HSUBI}
{\cal H}_I = -J \sum_{i=1}^{N-1} \sigma_i \sigma_{i+1}
+ h \sum_{i=2}^{N-1} \sigma_i - H [ \sigma_1 - \sigma_N ] \ .
\end{equation}

For $h>0$ the interface lies at $k=1$; for $h<0$ it unbinds to
$k=\infty$. $h=0$ is a multiphase point where every interface position
has the same energy. For classical spins, $S=\infty$,
the ground state  
and hence the multiphase point are maintained as the spin anisotropy
is decreased from $D=\infty$.

\begin{figure}
\centerline{\psfig{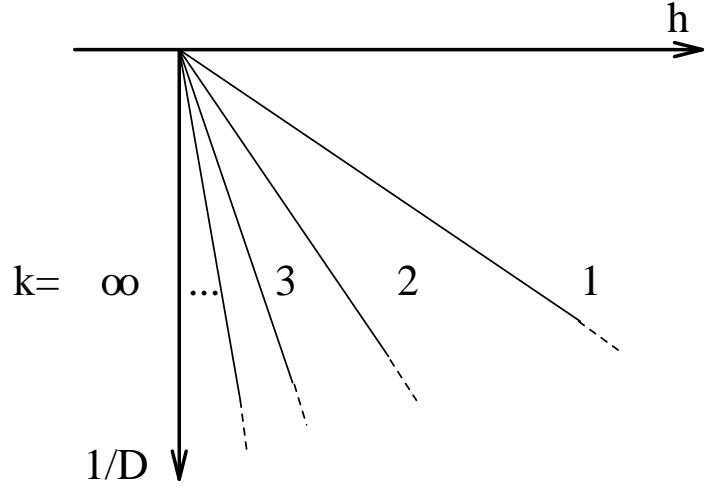}}
\vskip 0.5cm
\caption{Schematic representation of interface layering transitions in
the Heisenberg model with strong uniaxial spin anisotropy, $D$, as the
magnetic field $h$ changes sign. All interface phases $k$ appear in the phase diagram.}
\label{fig:pd}
\end{figure}

Our aim here is to study the way in which this degeneracy is lifted by
quantum fluctuations when
$D \gg J$ and
$S$ is large but finite. We find that the interface unbinds through
an infinite sequence of
first order transitions as $h \rightarrow 0^{+}$, as illustrated
schematically in Fig. \ref{fig:pd}. 
The existence of the transitions follows from considering the
structure of degenerate perturbation theory around the multiphase
point. To start the analysis we write the Hamiltonian~(\ref{HAMIL})
in bosonic form using the Dyson-Maleev transformation [\onlinecite{DM,SVM}]
\begin{eqnarray}
S_i^z & = & \sigma_i ( S - a_i^+ a_i) \nonumber \\
S_i^+ & = & \sqrt{2S} \left(
\delta_{\sigma_i,1} \left[ 1 - {a_i^+ a_i \over 2S} \right] a_i +
\delta_{\sigma_i,-1} a_i^+ \left[ 1-{a_i^+ a_i\over 2S} \right]
\right) \nonumber \\
S_i^- & = & \sqrt{2S} \left( \delta_{\sigma_i,1} a_i^+
+ \delta_{\sigma_i,-1} a_i \right) \ ,
\end{eqnarray}
where $\delta_{a,b}$ is unity if $a=b$ and is zero otherwise,
$a_i^+$ ($a_i$) creates (destroys) a spin excitation at
site $i$, and $\sigma_i$ specifies the sign of the $i$th spin.
The resulting Hamiltonian is
\begin{equation}
\label{HAM}
{\cal H} ( \{ \sigma_i \} ) = {\cal H}_I + {\cal H}_0
+ V_{||} + V_{\not{\parallel}} +V_4 \ ,
\end{equation}
where ${\cal H}_I$ is given in Eq. (\ref{HSUBI}), 
\begin{equation}
{\cal H}_0 = \sum_{i=2}^{N-1} \Biggl[  2D + J \sigma_{i}
 ( \sigma_{i-1} +
\sigma_{i+1} ) 
 - h \sigma_{i} ) \Biggr]
S^{-1} a_{i}^+ a_{i},
\label{eqn:h0}
\end{equation}
$V_4$ represents the four operator terms proportional to $1/S^2$, and
$V_{||}$ ($V_{\not{\parallel}}$) is the interaction between spins 
which are parallel (antiparallel)
\begin{equation}
V_{||} =-\sum_{i=2;i \ne k}^{N-1}JS^{-1} (a_{i}^+ a_{i+1} + a_{i+1}^+ a_{i} ),
\label{eqn:vpar}
\end{equation}
\begin{equation}
V_{\not{\parallel}} 
= -J S^{-1}(a_{k}^+ a_{k+1}^+ + a_{k+1} a_{k} ).
\label{eqn:vnotpar}
\end{equation}

We work to lowest order in $1/S$ and therefore neglect terms like
$V_4$ which are higher
order than quadratic in the boson operators. 

To understand the structure of the phase diagram near the multiphase
point it is most convenient to calculate the energy difference $\Delta
E_{k}=E_{k}-E_{k-1}$ where $E_k$ is the energy of the system with the
interface at position $k$ [\onlinecite{DuxY}]. In particular, contributions
to $E_{k}$ which 
are independent of $k$ do not affect the location of the interface and need
not be considered. The energies $E_k$ will be calculated at $h=0$ using
standard perturbation techniques [\onlinecite{Messiah}]
\begin{equation}
E_k = {}_k\!\langle 0| (V_{||} + V_{\not{\parallel}}) | 0 \rangle_k -
{}_k\!\langle 0| (V_{||} + V_{\not{\parallel}}) {Q_0 \over {\cal H}_0
- E_0} (V_{||} + V_{\not{\parallel}}) | 0 \rangle_k + \dots
\label{pert}
\end{equation}
where the vector $| 0 \rangle_k$ corresponds to the configuration with
 the interface at position $k$ and no excitation present and
$Q_0= 1 - | 0 \rangle_k\,{}_k\!\langle 0|$. All the
vectors $| 0 \rangle_k$ are eigenstates of ${\cal H}_0$ with the same
eigenvalue $E_0$. However, the perturbative term $(V_{||} +
V_{\not{\parallel}})$ conserves $\sum_i S_i^z$ and thus it can never
cause a transition between two different ground states.
Therefore we may use non-degenerate perturbation theory to
check whether the excitations can lift the degeneracy of the interface states.

\begin{figure}
\centerline{\psfig{figure=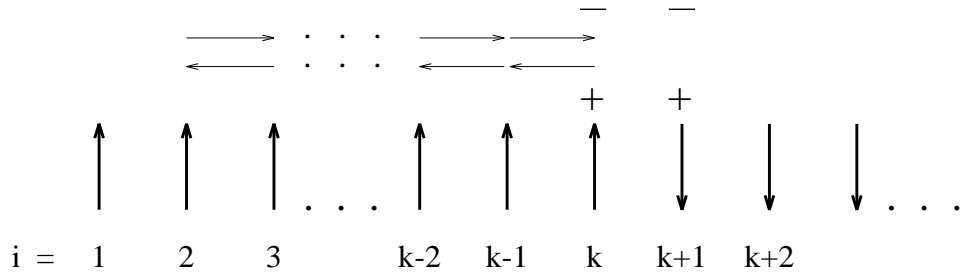,width=5.0in}}
\vskip 0.5cm
\caption{The process which gives the lowest order contribution to
the energy difference $\Delta E_k \equiv E_k-E_{k-1}$
between the interface at positions
$k$ and $k-1$. + (--) denotes the creation (destruction) of a spin
excitation by $V_{\not \parallel}$. An arrow is used to denote a hop
mediated by $V_{\parallel}$.
The process shown contributes to $E_k$ but not to $E_{k-1}$ because
the $i=1$ spin can not be flipped when $H \rightarrow \infty$.}
\label{fig:dek}
\end{figure}

Contributions to the energies $E_k$ arise from spin deviations
at the interface created by $V_{\not{\parallel}}$ which are propagated
away from and then back to the interface by $V_{||}$ and subsequently destroyed by 
$V_{\not{\parallel}}$. However only such processes which are
$k$-dependent are of interest to us.
The lowest order term which contributes to $\Delta E_{k}$
corresponds to an excitation which is created at the interface at
position $k$ and propagates to the surface and back before being
destroyed. This graph is illustrated in Fig. \ref{fig:dek}.  (This process
contributes to $E_{k}$, but does not occur for $E_{k-1}$.) It has a
contribution which follows immediately from $(2k)^{\rm th}$ order
perturbation theory as
\begin{equation}
\Delta E_k = - \frac{J^{2k}}{S(4D)^{2k-1}} + {\cal O}\bigl( {1 \over
D^{2k}} \bigr)
\end{equation}
\noindent where the terms in $J$ and $h$ in the denominator contribute
only to higher order in $1/D$.
$\Delta E_k$ is negative corresponding to a repulsive interaction
between the interface and the surface and hence as $h \rightarrow 0^{+}$
the interface unbinds through a series of first order phase
transitions with boundaries between the phases at
\begin{equation}
h_{k:k-1}=\frac{J^{2k}}{S(4D)^{2k-1}}.
\end{equation}
One feature of this calculation which  is seen again  for the ANNNI model
is the fact that the interface energy (here $\Delta E_k$)  involves  the
$(2k)$th power of the coupling constant, and not just the $k$th
power, as one might imagine for a classsical system [\onlinecite{DuxY}].  The point is
that the quantum fluctuation has to propagate from the interface to the
surface AND back.  As we will see later, this difference leads to
a crucial distinction between the way quantum fluctuations and
classical fluctuations lift the multiphase degeneracy for the ANNNI model.

\section{The ANNNI Model}

A similar formalism is now used to approach a more complicated problem:
the effect of quantum fluctuations where the multiphase point is a
point of infinite degeneracy for bulk rather than interface phases.
We take as our example 
the axial
next-nearest-neighbor
Ising or ANNNI model [\onlinecite{RJE}]. Rather
than a single wall interacting with a surface the phase structure is now
controlled by an infinite number of interacting walls
and we shall follow Fisher and Szpilka [\onlinecite{MEFXS}] in analyzing the phase
structure in terms of the interactions between the walls. A brief
account of this work has been published elsewhere [\onlinecite{HMY}]. 

The Hamiltonian we consider is
\begin{equation}
{\cal H} =
- {J_0 \over S^2} \sum_{i \langle jj'\rangle} {\bf S}_{i,j}
\cdot {\bf S}_{i,j'} - {J_1 \over S^2} \sum_{i,j} {\bf S}_{i,j} \cdot
{\bf S}_{i+1,j} 
+ {J_2 \over S^2} \sum_{i,j} {\bf S}_{i,j} \cdot {\bf S}_{i+2,j}
  -  {D \over S^2} \sum_{i,j} ([S_{i,j}^z]^2 -S^2) ,
\label{AHAMIL}
\end{equation}
where $i$ labels the planes of a cubic lattice perpendicular to
the $z$-direction and $j$ the position within the plane.  Also
$\langle jj'\rangle$ indicates a sum over pairs of nearest
neighbors in the same plane and ${\bf S}_{i,j}$ is a quantum
spin of magnitude $S$ at site $(i,j)$.  For $D= \infty$, only
the states $S_{i}^z=\sigma_iS$, where $\sigma_i = \pm 1$ are
relevant and ${\cal H}$ reduces to the 
ANNNI model [\onlinecite{RJE}].
\begin{equation}
{\cal H}_A  = 
- J_0 \sum_{i \langle jj'\rangle} \sigma_{i,j}
\sigma_{i,j'} - J_1 \sum_{i,j} \sigma_{i,j} \sigma_{i+1,j}
+ J_2 \sum_{i,j} \sigma_{i,j} \sigma_{i+2,j} .
\end{equation}
The ground state of the ANNNI model is ferromagnetic for
$\kappa \equiv J_2/J_1 < 1/2$ and an antiphase structure with
layers ordering in the sequence $\{ \sigma_i \} = \{ \dots
1, 1, -1 , -1, 1,1,-1, -1\dots \}$ for $\kappa > 1/2$.  
$\kappa=1/2$ is a multiphase point[\onlinecite{BAK}, \onlinecite{MEFWS}],
where the ground state is
infinitely degenerate with all possible configurations of
ferromagnetic and antiphase orderings having equal energy.
For classical spins, $S=\infty$, the ground state (and therefore
a multiphase point) is maintained as $D$ is reduced from
infinity.

To describe how the degeneracy is broken at the multiphase
point when $S$ is large, but not infinite, we define a notation
similar to that of Fisher and Selke[\onlinecite{MEFWS}] using
$\langle n_1, n_2, \dots n_m \rangle$ to denote a state consisting of
domains of parallel spins with alternate orientation 
whose widths repeat periodically the sequence $\{n_1, n_2, \dots n_m\}$.

As in the previous section we use the Dyson-Maleev [\onlinecite{DM,SVM}]
transformation to
recast the Hamiltonian~(\ref{AHAMIL}) into bosonic  form (working to lowest
order in $1/S$) with the result
\begin{equation}
\label{AHAM}
{\cal H} ( \{ \sigma_i \} ) = E_0 + {\cal H}_0
+ V_{||} + V_{\not{\parallel}}  \ ,
\end{equation}
where $E_0 \equiv {\cal H}_A$,
\begin{eqnarray}
{\cal H}_0 & = & \sum_{i,j} \Biggl[  2\tilde{D} + J_1 \sigma_{i,j}
( \sigma_{i-1,j} + \sigma_{i+1,j} ) - J_2 \sigma_{i,j} ( \sigma_{i-2,j}
+ \sigma_{i+2,j} ) \Biggr] S^{-1} a_{i,j}^+ a_{i,j} \nonumber \\
& \equiv & \sum_{i,j} E_{i,j} S^{-1} a_{i,j}^+ a_{i,j} \ ,
\label{eqn:calh0}
\end{eqnarray}
with $\tilde{D}=D+2 J_{0}$
and $V_{||}$ ($V_{\not{\parallel}}$) is the interactions between spins
which are parallel (antiparallel)
\begin{equation}
V_{||} =  {1 \over S} \sum_{i,j} \Biggl[ - J_1
X(i,i+1;j)
(a_{i,j}^+ a_{i+1,j} + a_{i+1,j}^+ a_{i,j} )
 + J_2
X(i,i+2;j)
(a_{i,j}^+ a_{i+2,j} + a_{i+2,j}^+ a_{i,j} ) \Biggr]
\label{eqn:vparallel}
\end{equation}
\begin{equation}
V_{\not{\parallel}} =  {1 \over S} \sum_{i,j} \Biggl[ - J_1
Y(i,i+1;j)
(a_{i,j}^+ a_{i+1,j}^+ + a_{i+1,j} a_{i,j} ) + J_2
Y(i,i+2;j)
(a_{i,j}^+ a_{i+2,j}^+ + a_{i+2,j} a_{i,j} ) \Biggr] \ ,
\label{eqn:vnotparallel}
\end{equation}
where $X(i,i';j)$ [$Y(i,i';j$)] is unity if spins $(i,j)$ and
$(i',j)$ are parallel [antiparallel] and is zero otherwise.
We do not consider quantum fluctuations
within a plane, since the phase diagram is determined by the
interplanar quantum couplings. Moreover we shall work to leading order
in $1/S$ , in which case four-operator terms can be 
neglected [\onlinecite{NEGLECT}].  Also we will continue to use
non-degenerate  perturbation theory, since the perturbative term
$(V_{||} + V_{\not{\parallel}})$ cannot connect states in which the
wall is at different locations, since such states have different
values of $\sum_iS_i^z$.

\begin{figure}
\centerline{\psfig{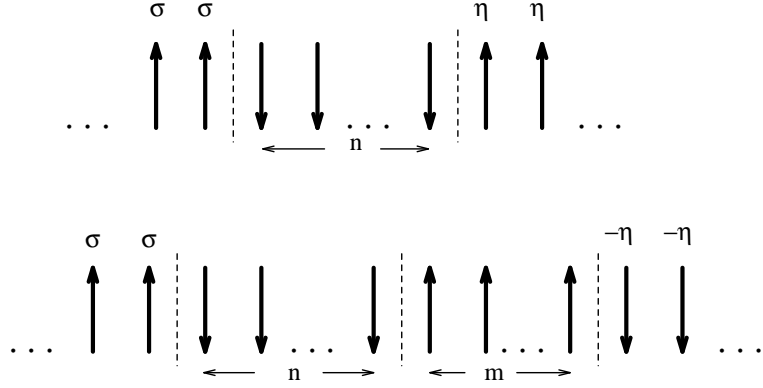}}
\vskip 0.5cm
\caption{Configurations needed to calculate the interaction energy for
two walls at separation $n$ (top) and three walls at separation $n$
and $m$ (bottom). When $\sigma=+1$ ($\eta=+1$) the left-most
(right-most) wall is positioned as shown. When $\sigma=-1$ ($\eta=-1$)
the left-most (right-most) wall does not exist.}
\label{fig:2w}
\end{figure}

The structure of the phase diagram will be constructed by considering in turn
$E_w$, the energy of an isolated wall; $V_2(n)$, the interaction energy
of two walls separated by $n$ sites; and generally
$V_k(n_1 , n_2 , \dots n_{k-1})$, the interaction energy of $k$ walls
with successive separations $n_1$, $n_2$, ... $n_{k-1}$ [\onlinecite{MEFXS}].
In terms of these quantities one may write the total energy
of the system when there are $n_w$ walls at positions $m_i$ as
\begin{eqnarray}
E = && E_0 + n_w E_w + \sum_i V_2(m_{i+1}-m_i)
 +  \sum_i
V_3(m_{i+2}-m_{i+1},m_{i+1}-m_i)\nonumber \\ 
& + & \sum_i
V_4(m_{i+3}-m_{i+2},m_{i+2}-m_{i+1},m_{i+1}-m_i)
 +  \dots ,
\end{eqnarray}
where $E_0$ is the energy with no walls present.
The scheme of Ref. [\onlinecite{RBG}] for
calculating the general wall potentials $V_k$ is illustrated in
Fig. \ref{fig:2w}.
Let all spins to the left of the first wall have $\sigma_i=\sigma$
and those to the right of the last wall have $\sigma_i=\eta$ for
$k$ even and $\sigma_i=-\eta$ for $k$ odd.  The energy of such a
configuration is denoted $E_k(\sigma, \eta)$.  If $\sigma=-1$
($\eta=-1$) the left (right) wall is absent.  Thus the energy
ascribed to the existence of $k$ walls is given by[\onlinecite{RBG}]
\begin{equation}
V_k(n_1, n_2, \dots n_{k-1}) = \sum_{\sigma , \eta = \pm 1}
\sigma \eta E_k (\sigma , \eta) \ .
\label{CONN}
\end{equation}
Contributions to $E_k$ which are independent of $\sigma$ or $\eta$
do not influence $V_k$.  
$E_{k}(\sigma,\eta)$ is calculated by developing the energy in powers
of the perturbations $V_{\not{\parallel}}$ which allows creation
(and annihilation) of a
pair of excitations straddling a wall and $V_{||}$ which allows the
excitations to hop within domains. We consider contributions to the
wall energy and to two- and three-wall interactions in turn.

\subsection{Wall energy}

\begin{figure}
\centerline{\psfig{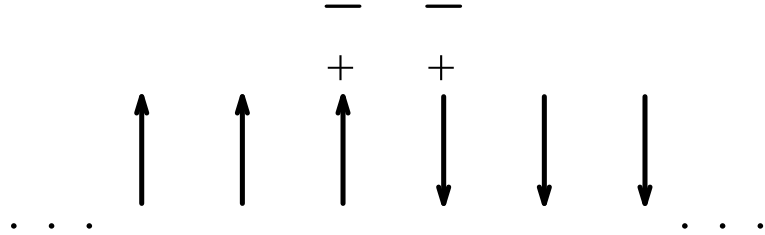}}
\vskip 0.5cm
\caption{The contribution from second order perturbation theory which
renormalizes the wall energy. + (--) denotes the creation (destruction)
of a spin excitation by $V_{\not{\parallel}}$.}
\label{fig:2pt}
\end{figure}

Contributions to the wall energy to second order in perturbation
theory arise from excitations which are created at a wall and then
immediately destroyed as shown in Fig. \ref{fig:2pt}. These effectively count the
number of walls and therefore lead to a renormalization of the wall
energy of 
\begin{equation}
E_{w}=2J_{1}-4J_{2}-\frac{J_1^2-2J_2^2}{4 \tilde{D} S}+{\cal O} \left(
\frac{J^3}{\tilde{D}^2 S} \right)
\label{eqn:Ewall}
\end{equation}
but since we work to leading order in $S^{-1}$, the $S^{-1}$
correction to $E_w$ will not affect the results for  $V_k$.

\subsection{Pair interactions}

\begin{figure}
\centerline{\psfig{figure=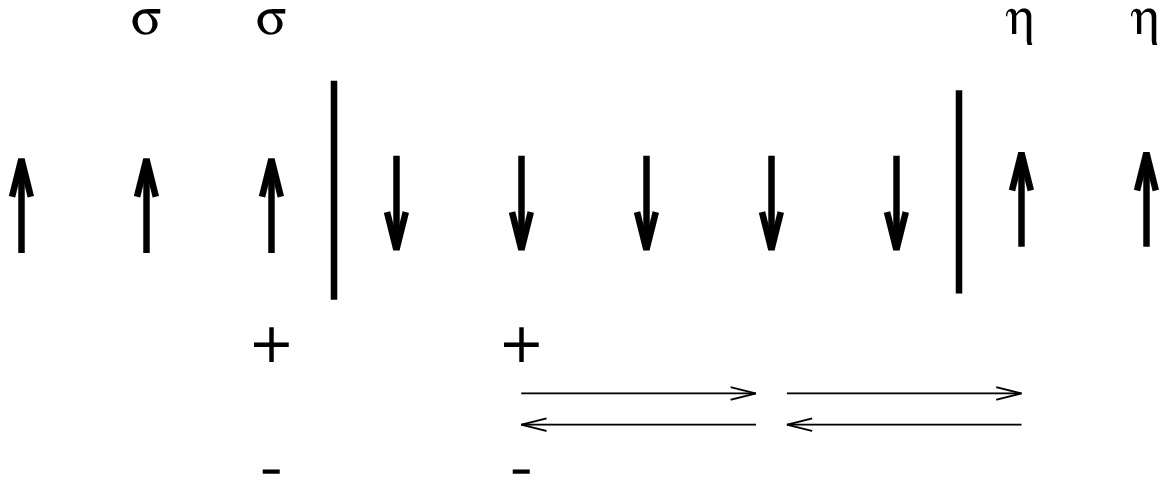}}
\vskip 0.5cm
\caption{Contribution to the two-wall interaction $V_2(n)$ for
$n=5$ in analogy with the unbinding problem of Fig. \protect\ref{fig:dek}.
This process contributes to $V_2(5)$ at order $J_2^6/D^5$.}
\label{fig:2w5}
\end{figure}

\begin{figure}
\centerline{\psfig{figure=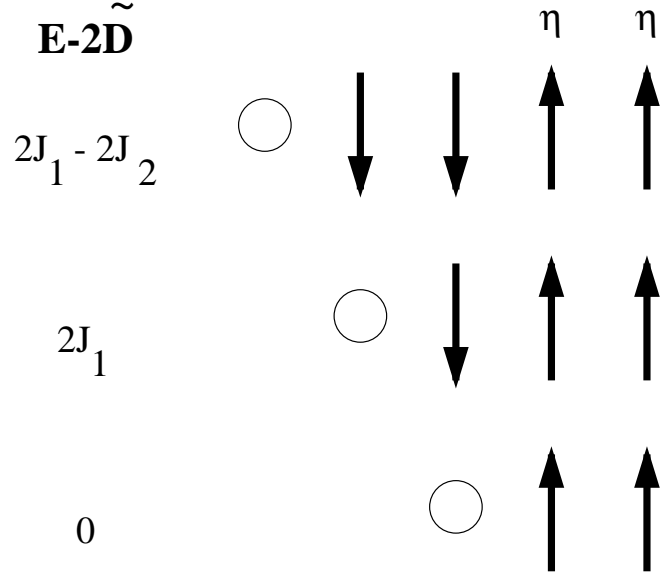}}
\vskip 0.5cm
\caption{The energy, $E$, of an excitation as a function of position
near a wall.
We give $E-2 \tilde D$ when the excitation is created at the circled
site.  Thus when the excitation is on a site next nearest neighboring
to the wall its energy is $e_1=2 \tilde D + 2J_1 - J_2 (1 - \eta)$
and when nearest neighboring its energy is 
$e_2=2 \tilde D + (J_1-J_2)(1-\eta)$.
When the wall is absent, these formula give
the correct energy of an excitation when it is not near a wall.
Thus $de_1/d\eta =J_2$ and $de_2/d\eta =J_2-J_1$.}
\label{fig:z}
\end{figure}

\begin{figure}
\centerline{\psfig{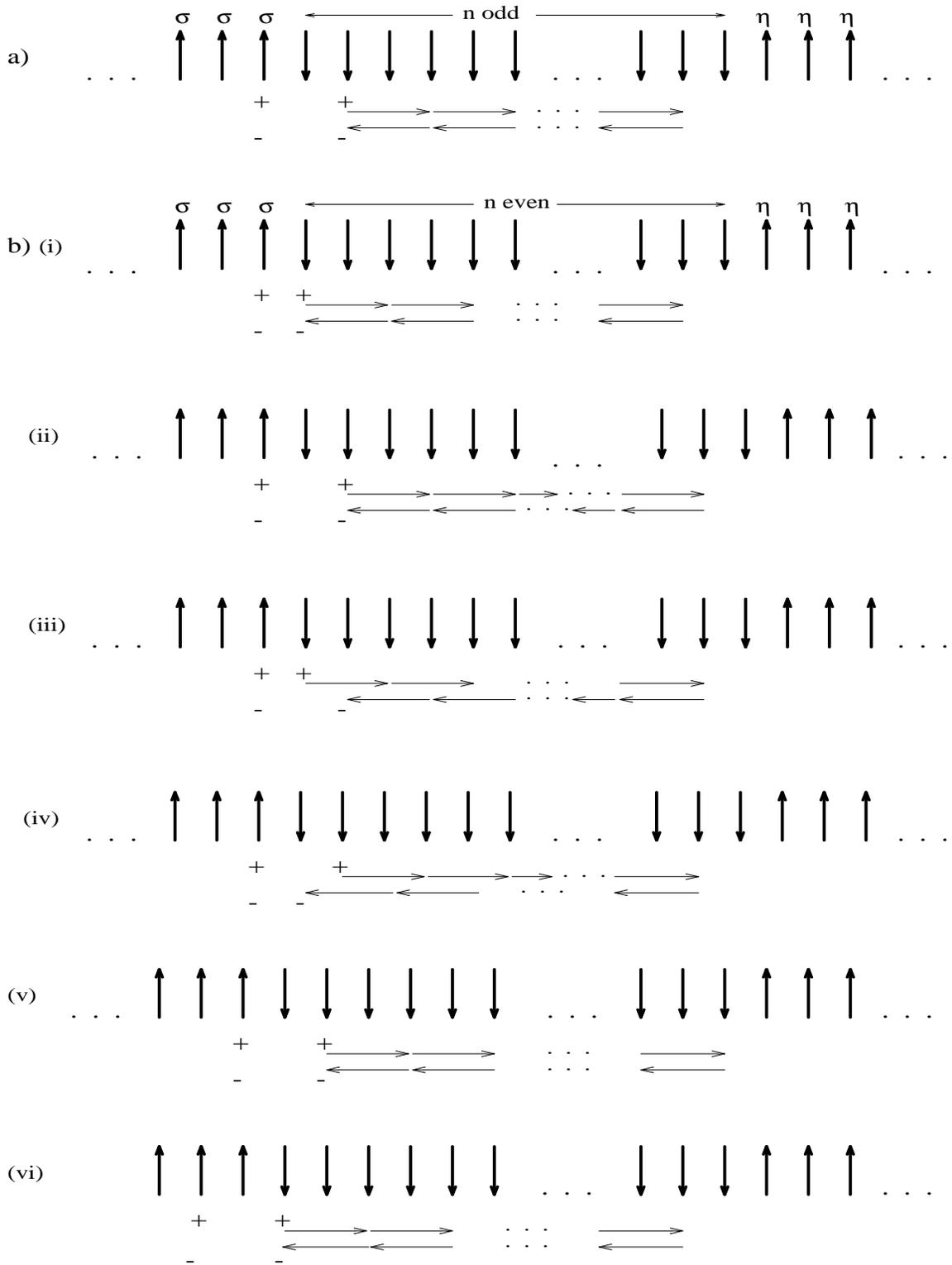}}
\vskip 0.5cm
\caption{Excitations which contribute to the 2-wall interaction
$V_2(n)$ for  (a) $n$ odd and (b) $n$ even. + (--) denotes the creation
(destruction) of a spin excitation by $V_{\not{\parallel}}$. An arrow
denotes a hop mediated by $V_{\parallel}$.}
\label{fig:2wnoe}
\end{figure}

The lowest order contributions to $V_2(n)$ are obtained by creating an
excitation at, say, the left wall using $V_{\not{\parallel}}$ and then
using $V_{||}$ for it to hop to the right wall and back. Because we
assume the existence of the left wall, this contribution implicitly
includes a factor $\delta_{\sigma,1}$.  Now we look for the
lowest-order (in $J/D$) contribution which also has a dependence on
$\eta$. In analogy with the unbinding problem, we might
consider processes in which the excitation hops beyond the wall.
Since such a term can not occur when the wall is actually
present, it will carry a factor
$\delta_{\eta,-1}$.
For $n$ odd, 
we illustrate this process in Fig. \ref{fig:2w5}, and see that it
gives a contribution to $V_2(n)$ of order $J_2^{n+1}/D^n$.  As we
shall see, there is actually a slightly different process which
comes in at one order lower in $J/D$.  To sense the
presence of the right-hand wall, note that $E_{i,j}$ in
Eq. (\ref{eqn:calh0}) will depend on $\eta$ if the $i$ is within
two sites of the wall.  Therefore it is only necessary to
hop to within two sites of the right wall, as shown in
Fig. \ref{fig:z}, for an energy
denominator $({\cal H}_0 -E_0)$ in the series expansion
(\ref{pert}) to depend on $\eta$. This process is of lower
order in $J/D$ because it takes two interactions to hop back
and forth but only one to sense the potential
via an energy denominator.
Accordingly, in
contrast to the interface unbinding considered in Sec. 2, it is
necessary to retain the terms in the $J$'s in the energy
denominators to obtain the leading order contribution to
$V_2(n)$. We consider separately $n$ odd and $n$ even. \\
\underline{$n$ odd}\\
To lowest order the processes which contribute are those shown in
Fig. \ref{fig:2wnoe}a. For a domain of $n$ spins with $\sigma_i=-1$,
$(n-1)^{\rm th}$ order perturbation theory gives
\begin{eqnarray}
\label{odd}
E_2(\sigma,\eta) =  &&  2\delta_{\sigma,1} J_2^{n-1}
S^{-1}(-1)^{n-2}\{4 \tilde{D}+2J_1\}^{-2}\{4 \tilde{D} +2J_1-2J_2\}^{-(n-5)}
\nonumber \\ && \times \{ 4 \tilde D +2J_1 - J_2 (1-\eta) \}^{-1} \ .
\end{eqnarray}
In writing this result we dropped all lower-order terms because
they do not depend on both $\sigma$ and $\eta$. Here and below, 
the dependence on $\sigma$ is contained in the factor $\delta_{\sigma,1}$
because we assume the existence of the left-hand wall.
 The
energy denominators are constructed as follows.  The left-hand
excitation has energy $2 \tilde D$ since it is next to a wall.
The right-hand excitation has the energy according to its position
as illustrated in Fig. \ref{fig:z}.  The prefactor of 2 arises because the
initial excitation can be near 
either wall and the overall factor $(-1)^{n-2}$ arises from the
$(-1)$ associated with each energy denominator. 
 Adding the contributions from~(\ref{odd})
appropriately weighted as in~(\ref{CONN}) gives
\begin{eqnarray}
V_{2}(n)&=& {2 J_2^{n-1} S^{-1} (-1)^{n-2} \over \{ 4 \tilde{D} + 2
J_1\}^2 \{ 4 \tilde{D} + 2 J_1 - 2 J_2 \}^{n-5}} \biggl \{ {1 \over 4
\tilde{D} + 2 J_1} - {1 \over 4 \tilde{D} +2 J_1 -2 J_2} \biggr\} \label{oddd} \\
& & \nonumber \\
& & \hskip 5.0cm =4J_2^{n} S^{-1}/ (4\tilde{D})^{n-1} + {
\cal O}(1/ \tilde{D}^n),\;\;\;\;\;\;\;\mbox{$n$ odd.} 
\label{V2odd}
\end{eqnarray}

\noindent
Note that there is no term ${\cal O}(1/\tilde{D}^{n-2})$. This
is because to this order the energy denominators are independent of
the $J$'s. Hence to this order $E_k(\sigma,\eta)$ is independent of $\eta$
and the sum in
Eq.~(\ref{CONN})
is zero. Similarly terms
from $n^{th}$ order perturbation theory (in which one $J_2$ hop is
replaced by two $J_1$ hops) do not contribute ${\cal O}(1/\tilde{D}^{n-1})$.

\noindent \underline{$n$ even}\\
For even $n$ several diagrams contribute to leading order, i.~e., at
$n$th order perturbation theory. These are shown in Fig. \ref{fig:2wnoe}b.
As an example we give the contributions to the energy from the diagram
(b)(iii).  Again we drop all terms which do not depend on both
$\sigma$ and $\eta$.  Thus

\begin{eqnarray}
E_2^{({\rm iii})}(\sigma , \eta) = && 2 (-1)^{n-1} \delta_{\sigma ,1}
\left( {n-2 \over 2 } \right) J_1^2 J_2^{n-2} S^{-1}
(4 \tilde D)^{-1} (4 \tilde D + 2 J_1)^{-1} \nonumber \\
&& (4 \tilde D + 2J_1 - 2J_2)^{-(n-4)} [ 4 \tilde D + 2J_1 - J_2 (1-
\eta) ]^{-1} \ ,
\label{even}
\end{eqnarray}
where the superscript iii indicates a contribution from diagram iii of
Fig. \ref{fig:2wnoe}, the prefactor 2 comes from including the
contribution of the
mirror image diagram, the prefactor $(-1)^{n-1}$ is the sign of
$n$th order perturbation theory,
the factor $(n-2)/2$ is the number of places the single ($J_1$)
hop can be put, and
$\delta_{\sigma,1}$ indicates
that this contribution assumes the existence of the left-hand wall.
To leading order in $\tilde D$, the $\eta$-dependence is contained in
\begin{eqnarray}
E_2^{({\rm iii})} (\sigma \eta) & = & (-1)^{n-1}(n-2) \delta_{\sigma,1}
J_1^2 J_2^{n-2}S^{-1} (4 \tilde D)^{-(n-2)}
(4 \tilde D +  \eta de_2 / d \eta )^{-1} \nonumber \\
& \approx & (-1)^n(n-2) \eta \delta_{\sigma,1}
J_1^2 J_2^{n-2}S^{-1} (4 \tilde D)^{-n} ( d e_2 /d \eta) \ .
\end{eqnarray}
Using $de_2/d \eta = J_2$, we get
\begin{equation}
V_2^{({\rm iii})} = 2 (n-2)J_1^2 J_2^{n-1}S^{-1} (4 \tilde D)^{-n} \ .
\end{equation}
We treat the other diagrams of Fig. \ref{fig:2wnoe} similarly.  Dropping terms which
do not depend on both $\sigma$ and $\eta$ and working to lowest order
in $(\tilde D)^{-1}$, we get
\begin{eqnarray}
E_2 ( \sigma , \eta ) &=& \eta \delta_{\sigma ,1} J_2^{n-2} S^{-1}
(4 \tilde D)^{-n} \Biggl[ 2 J_1^2 (de_2/d \eta)
+ {1 \over 2} J_1^2 (n-2)^2 (de_2/d \eta) \nonumber \\
&& \ \  + (n-2) J_1^2 (de_2/d \eta)
+ (n-2) J_1^2 (de_2/d \eta)
+ 2J_2^2 (de_1/d \eta)
+ 2J_2^2 (de_2/d \eta) \Biggr] \ ,
\end{eqnarray}
where the contributions are from each diagram of Fig \ref{fig:2wnoe}, written in the
order in which they appear in the figure.  Thus for $n$ even we have
\begin{eqnarray}
V_2(n) & = & S^{-1} (4 \tilde D)^{-n} J_2^{n-2} \Biggl[
4J_1^2 J_2 + J_1^2J_2 (n-2)^2
+ 4 (n-2)J_1^2J_2 + 4 J_2^2(J_2-J_1) + 4J_2^ 3 \Biggr]
\nonumber \\ &=&\frac{J_2^{n-1}}{(4 \tilde{D})^n S}(n^2 J_1^2 -4J_1J_2+8J_2^2),
\;\;\;\;\;\; \mbox{$n$ even} ,
\label{V2even}
\end{eqnarray}
where we used $de_2/d\eta=J_2$ and $de_1/d\eta=J_2-J_1$.

\begin{figure}
\centerline{\psfig{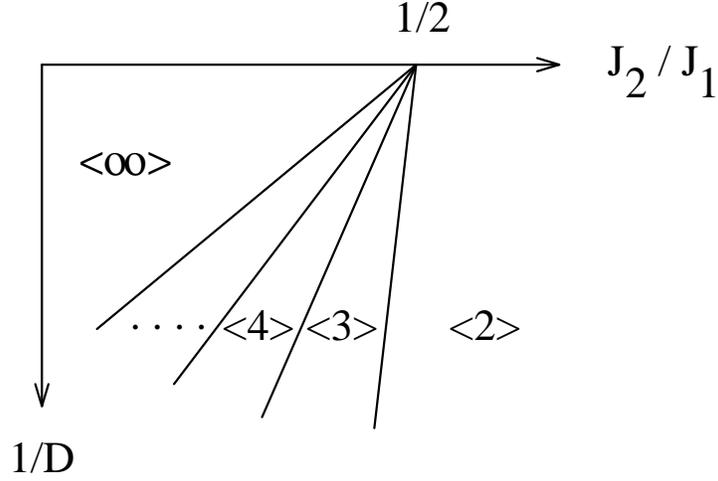}}
\vskip 0.5cm
\caption{Schematic phase diagram of the Heisenberg version of the 
ANNNI model with strong uniaxial spin anisotropy $D$ showing
the effect of quantum fluctuations.}
\label{fig:pda}
\end{figure}

Fisher and Szpilka [\onlinecite{MEFXS}] have shown that the phase
sequences can be determined graphically by constructing the lower
convex envelope of $V_2(n)$
versus $n$. The points $[n,V_2(n)]$ which lie on the envelope
correspond to stable phases. The pair interactions defined by the
expressions (\ref{V2odd}) and (\ref{V2even}) correspond already to a convex
function for $n << (\tilde{D}/J)^{1/2}$. Hence, in this regime, we
expect within the two-wall approximation a sequence
of phases $\langle 2 \rangle, \langle 3 \rangle,
\langle 4 \rangle, \ldots$ as
shown schematically in Fig. \ref{fig:pda}.
The widths of the phases $\langle n \rangle$
can be estimated using the fact that each phase is stable over an
interval
$\Delta E_w = n [V_2(n-1) - 2 V_2(n) - V_2(n+1) ]$
[\onlinecite{MEFXS}].  Therefore, using (\ref{eqn:Ewall}) we can say that
the width $\Delta(J_2/J_1)$ occupied by the phase $\langle n \rangle$ in
Fig. \ref{fig:pda} is ${\cal O}((J_2/D)^{n-1})$ for $n$ odd and\ ${\cal
O}((J_2/D)^{n-2})$ for $n$ even. 
This sequence of layering through unitary steps $\langle n \rangle \to
\langle n+1 \rangle$ will not be obeyed for large $n$, i. e., for
$n \sim (\tilde{D}/J)^{1/2}$, because then $V_2(n)$ will suffer from
strong even-odd 
oscillations. Moreover, for large $n$, the entropy of more
complicated perturbations may dominate the physics. A discussion of
this is given in Sec. 5. Here we go on to consider the effect of
3-wall interactions which can split the phase boundaries $\langle n
\rangle : \langle n+1 \rangle$ where there is still a
multiphase degeneracy of all states comprising domains of length $n$
and $n+1$.

\section{Three-wall interactions}
\begin{figure}
\centerline{\psfig{figure=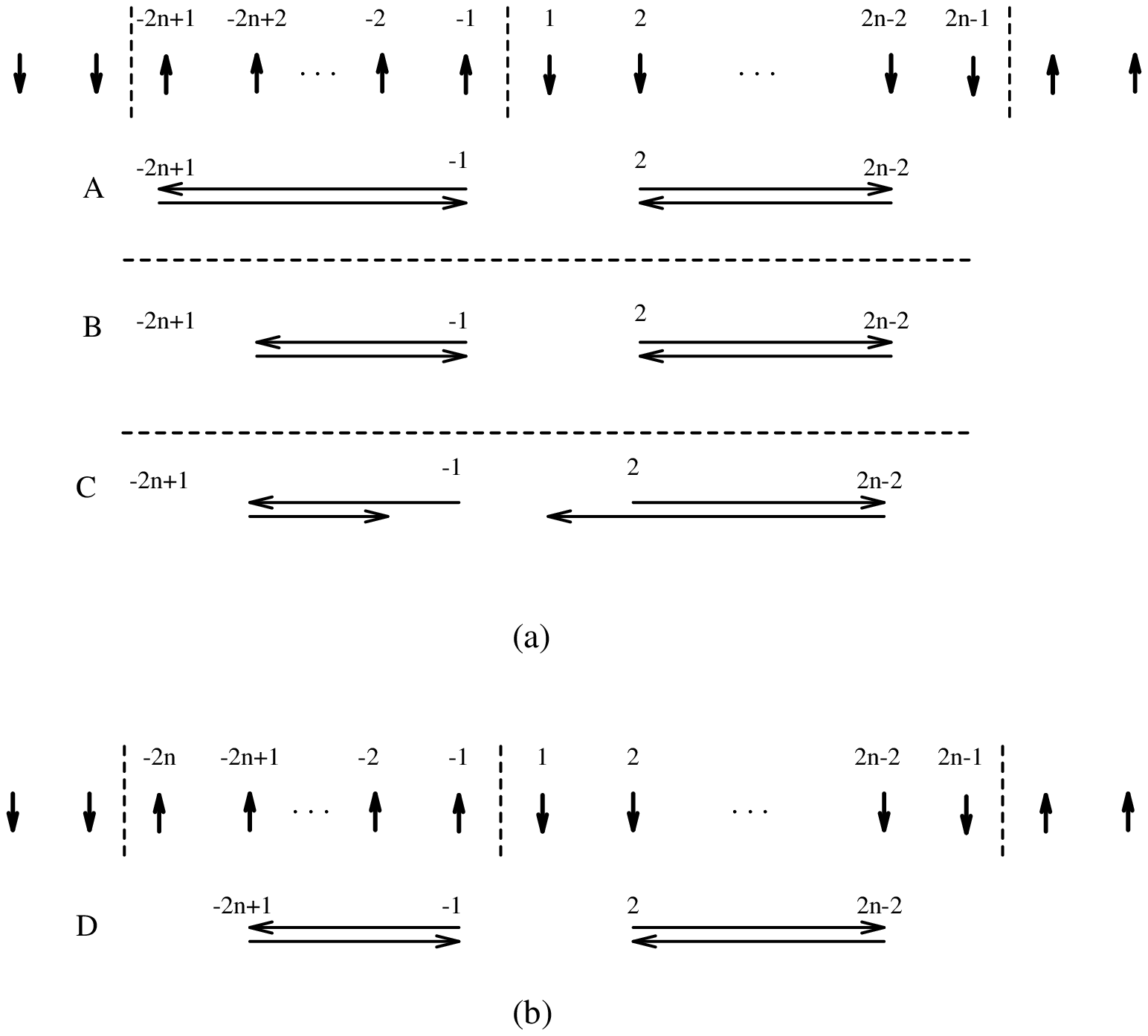,height=4.5in}}
\vskip 0.5cm
\caption{The diagrams needed to calculate $F(2n-1,2n)$ to leading
order: contributions to (a) $V_3(2n-1,2n-1)$ (b) $V_3(2n,2n+1)$.}
\label{fig:f1}
\end{figure}

Three-wall interactions are needed to analyze the stability of the  
$\langle n
\rangle : \langle n+1 \rangle$ phase boundary to mixed
phases of $\langle n \rangle$ and $\langle n+1
\rangle$.
The condition that the boundary be stable is [\onlinecite{MEFXS}]
\begin{equation}
\label{FEQ}
F(n,n+1) \equiv V_3(n,n)-2V_3(n,n+1)+V_3(n+1,n+1)<0.
\label{eqn:F}
\end{equation}
\noindent
Consider first the calculation of $F(2n-1,2n)$. 
The diagrams which contribute in leading order to $V_3(2n-1,2n-1)$ and
$V_3(2n,2n-1)$  are shown in Figs. \ref{fig:f1}a
and \ref{fig:f1}b, respectively. To leading order in $1 / \tilde{D}$,
$V_3(n+1,n+1)$ does not contribute to $F(n,n+1)$. 
Figure \ref{fig:f1} aims to emphasize the positions of the
initial excitation and the closest approaches to the neighboring
domain walls. One must also consider the position of the first
neighbor hops in B and C and the sequence of the hops when calculating
the contribution of the diagrams.

An explicit calculation of the contributions of the relevant diagrams
would be extremely tedious. However what concerns us here is the
sign of $F(2n-1,2n)$.  If $N_i$ is the contribution to $F$ of diagrams
of type $i$ in Fig. \ref{fig:f1} ,
\begin{equation}
F(2n-1,2n)=2 N_{\rm A} + 2 N_{\rm B} + 2 N_{\rm C} - 2 N_{\rm D}, 
\end{equation}
where the factors of $2$ multiplying $N_{\rm A}$,  $N_{\rm B}$ and
$N_{\rm C}$ account for the mirror image diagrams and that multiplying
$N_{\rm D}$ occurs because of the 2 in Eq. (\ref{FEQ}).

We shall now show that $F(2n-1,2n)<0$. Consider a diagram in which the
hops occur in the same order in A, B, C and D and the $J_1$ hops in B
and C are, say, nearest the outer walls. The matrix elements $m_i$ of all
types of diagram carry a negative common factor (the sign arising
because we are considering even-order perturbation theory) and their
ratios are $m_{\rm A}/m_{\rm D}=1$ and
$m_{\rm A}/m_{\rm B}=m_{\rm A}/m_{\rm C}=J_2^2/J_1^2$.

We must also expand the difference in the energy denominators in a way
analogous to the step between equation (\ref{oddd}) and (\ref{V2odd}),
but here to second order in $J/\tilde{D}$. Using (\ref{CONN}), the
contribution of each diagram to the appropriate $N_i$ may be written
\begin{eqnarray}
& & \sum_{\sigma,\eta} \sigma \eta \biggl[{m_i\over (4\tilde{D})^{4n-5}S}\left(f_1+{f_2+f_3\sigma +f_4 \eta \over
(4\tilde{D})} 
+{f_5+f_6\sigma+f_7\eta+f_8\sigma^2+f_9\eta^2+f_{10}\sigma\eta
\over (4\tilde{D})^2}+\ldots \right) \biggr] \nonumber \\
& & \nonumber \\
& & \hskip 7.0cm = {4 m_i f_{10} \over
(4\tilde{D})^{4n-3}S} + {\cal O}({1  \over (4\tilde{D})^{4n-2}})
\label{eqn:fs}
\end{eqnarray}

\noindent where the coefficients $f$ depend only
on $J_1$ and $J_2$. When the sum is taken only the term $f_{10}$
multiplying $\sigma \eta$ survives. For
diagrams of type A, $f_{10}$ is $J_2(J_2-J_1)$, while for B, C and
D it is $J_2^2$. Therefore these
diagrams give a contribution to $F$ proportional to
\begin{equation}
-J_2^2J_1(2 J_1-J_2) <0 \ .
\end{equation}
\noindent 
The contributions to $F$ of the other diagrams in B and C
(which correspond to a different position of the
first neighbor hop) is proportional to $-J_1^2 J_2^2$.
Hence $F(2n-1,2n)<0$ and the $\langle 2n-1 \rangle:
\langle 2n \rangle$ boundaries are stable.

\begin{figure}
\centerline{\psfig{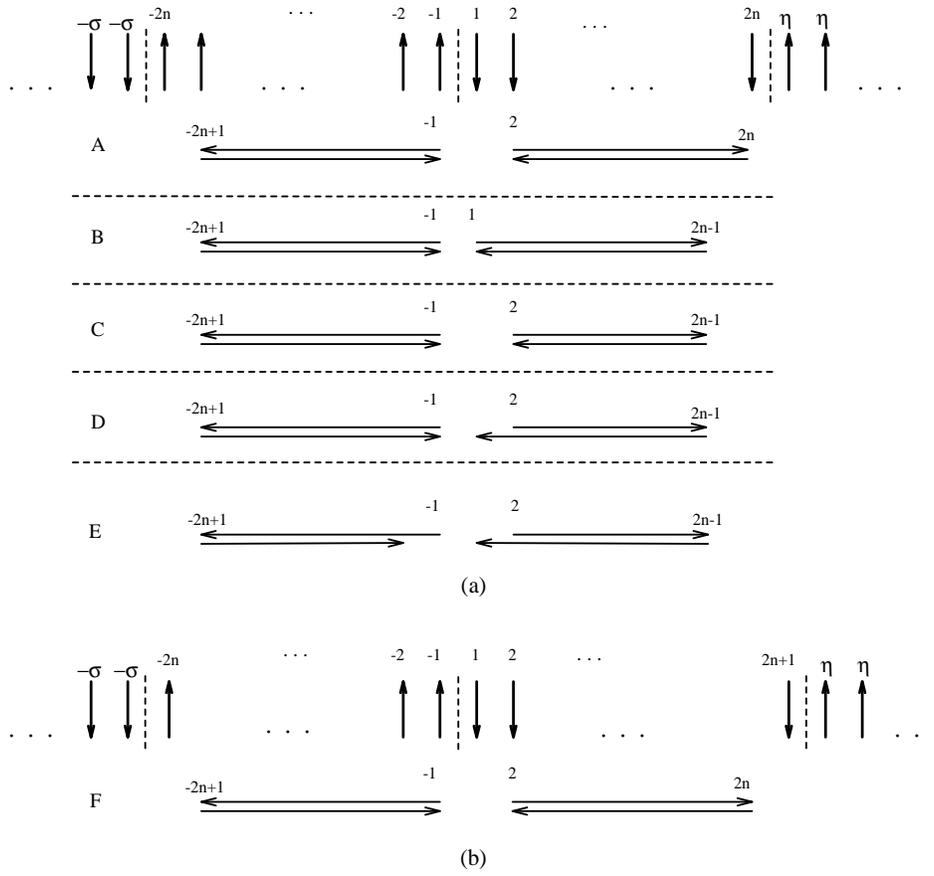}}
\vskip 0.5cm
\caption{The diagrams needed to calculate $F(2n,2n+1)$ to leading
order: contributions to (a) $V_3(2n,2n)$ (b) $V_3(2n,2n+1)$.}
\label{fig:f2}
\end{figure}

A similar argument holds for $F(2n,2n+1)$ for $n>1$. The relevant
diagrams are shown in Fig. \ref{fig:f2}. They contribute
\begin{equation}
F(2n,2n+1)=2 N_{\rm A} + 2 N_{\rm B} + 2 N_{\rm C} + 2 N_{\rm D} +2
N_{\rm E} -2 N_{\rm F}.
\end{equation}
Using the same argument as above
\begin{equation}
N_{\rm A} + N_{\rm B} - N_{\rm F} \propto -J_2^2J_1(J_1-J_2)<0.
\end{equation}
$N_{\rm C}$, $N_{\rm D}$, $N_{\rm E}$, and the other orderings of
$N_{\rm B}$ are negative and hence $F(2n,2n+1)<0$.  Thus
the phase boundaries $\langle 2n \rangle : \langle 2n+1 \rangle$
are first order for $n > 1$.

For the $\langle 2 \rangle : \langle 3 \rangle $ boundary 
different diagrams contribute to $F(2,3)$.
Indeed the second order
expansion of the
energy denominators [as in Eq. (\ref{eqn:fs})] gives a zero
contribution.
Accordingly, the calculation of $F(2,3)$ requires
going to higher order in $(J_2/\tilde D)$.  This calculation is
carried out in detail in Appendix A and shows that the
$\langle 2 \rangle : \langle 3 \rangle$ boundary is also stable.

\section{Large $ \lowercase{n}$ analysis}

For small $n$, we have seen that the leading contribution to
$V_2(n)$ is of order $D(J_2/D)^x/S$, where the value of $x$
corresponds to the minimum number of steps needed to go from
near one wall to near the other one and back: $x = 2 [n/2]+1$,
where $[x]$ is the integer part of $x$.  As $n$ increases, the
contributions
from longer paths, although individually less important, can
become dominant because of their greater entropy.  To allow
for this possibility we now carry out perturbation theory in
terms of the exact eigenstates for one excitation
in each block of parallel spins.  In this formulation,
the unperturbed Hamiltonian is the sum of the Hamiltonians
of each domain of parallel spins when all interactions with
neighboring domains are removed. Thus from equations (\ref{eqn:calh0})
and (\ref{eqn:vparallel}) the unperturbed Hamiltonian for a block of
parallel spins from sites $I$ to $J$ inclusive can be written 
\begin{equation}
{\cal H}_0^{(I,J)} = {\sum_{i,j}}^\prime
J_{ij} S^{-1} ( a_i^+ - a_j^+)(a_i - a_j) + \sum_i 2
\tilde DS^{-1} a_i^+ a_i \ ,
\end{equation}
where $J_{i,j}= J_1 \delta_{j,i+1} -J_2 \delta_{j,i+2}$ 
 and the prime on the summation indicates that the sum is restricted so
that both indices are actually in the block.  The matrix representation
of ${\cal H}_0^{(I,J)}$ is given explicitly in Appendix B.   It
follows from (\ref{eqn:h0}) and (\ref{eqn:vnotparallel}) that  the
perturbation $V^{(s)}$, associated with a wall between sites $s$ and
$s+1$ is
\begin{equation}
\label{VWALL}
V^{(s)} = W^{(s)} + X^{(s)}+Y^{(s)}  
\end{equation}
where
\begin{eqnarray}
W^{(s)} & = & J_2 S^{-1} \Bigl( a_{s-1}^+ a_{s-1} + a_s^+ a_s
+ a_{s+1}^+ a_{s+1} + a_{s+2}^+ a_{s+2} \Bigr)
- J_1S^{-1} \Bigl( a_s^+ a_s + a_{s+1}^+ a_{s+1} \Bigr)
\nonumber \\ &\equiv& S^{-1} \sum_k W_k^{(s)} a_k^+ a_k  \ ,
\label{eqn:ws}
\end{eqnarray}
\begin{equation}
X^{(s)} = J_2S^{-1} \Bigl( a_{s+1}^+ a_{s-1}^+
+ a_{s+2}^+ a_s^+ \Bigr) - J_1 S^{-1} \Bigl(
a_{s+1}^+ a_s^+ \Bigr) \equiv \sum_{i<j}
S^{-1} X_{ij}^{(s)} a_i^+ a_j^+ \ ,
\end{equation}
and $Y^{(s)}=\Bigl( X^{(s)} \Bigr)^+$.

In this formulation  it is not natural to  calculate $V_2(n)$
directly.  Instead one calculates the total energy of  given  configurations
from which $V_2(n)$ is easily deduced.  We start by calculating
the total energy of a configuration with a single wall between sites
$0$ and $1$. This gives  the wall energy  as
\begin{equation}
\label{WALL1}
E_1^{(2)} = \langle 0 | Y^{(0)} {Q_0 \over {\cal E}} X^{(0)}
| 0 \rangle = S^{-2} \sum_{ijkl} X^{(0)}_{kl} X^{(0)}_{ij}
\langle 0| a_k a_l {Q_0 \over {\cal E}} a_i^+
a_j^+ | 0 \rangle 
\label{eqn:e12}
\end{equation}
where ${\cal E} = E_0 - {\cal H}_0$,  with $E_0$  the ground state
energy, defined to be zero in this context.
Here we  have introduced  the notation that the subscript on $E$ specifies
the number of walls, the superscript the order in perturbation
theory, and the  arguments (if any) the separations between walls.

To evaluate  (\ref{eqn:e12})  and similar expressions we now introduce the
exact eigenstates for a single excitation on either side
of the wall when interactions across the wall are ignored.  For
a block of parallel spins occupying sites $I$ through $J$,
inclusive, these single-particle eigenstates satisfy
\begin{equation}
\label{EIGEQ}
\sum_j \Bigl( {\cal H}_0^{(I,J)} \Bigr) _{ij}
\phi_\alpha^{(I,J)}(j) = S^{-1} \epsilon_\alpha^{(I,J)}
\phi_\alpha^{(I,J)} (i) \ .
\end{equation}
Later we write $\epsilon^{(I,J)} \rightarrow \epsilon^{(J-I+1)}$.

\begin{figure}
\centerline{\psfig{figure=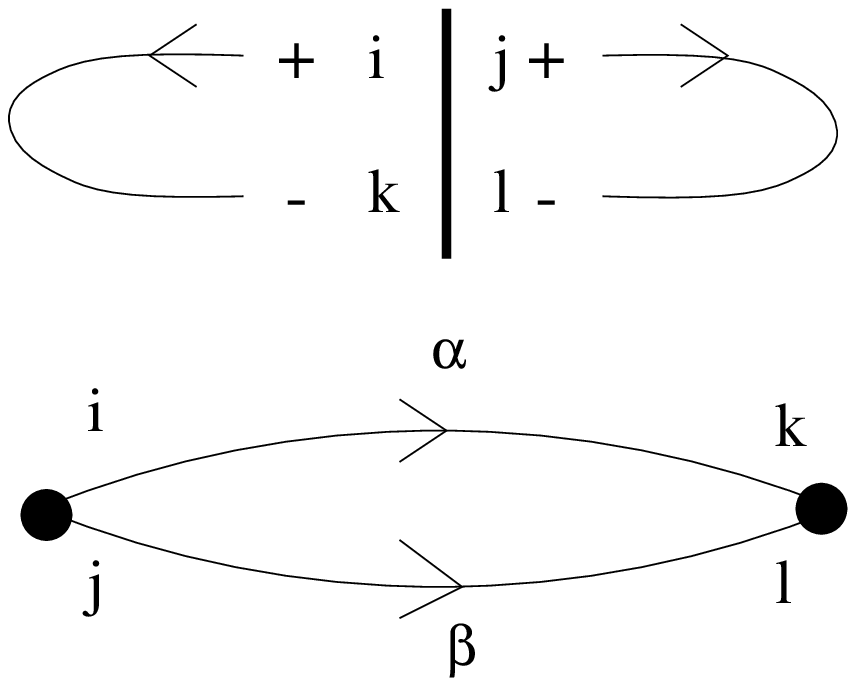,width=4in,height=2.5in}}
\vspace{0.5cm}

\caption{Contribution to $E_1^{(2)}$ in second-order
perturbation theory. Top: real space representation showing
sites $i$ and $j$ near the wall where the excitations are
created and sites $k$ and $l$ where they are destroyed.
Bottom: Feynmann diagram representation.}
\label{fig:e12}
\end{figure}

To evaluate Eq. (\ref{WALL1}) in terms of the exact eigenstates
notice that $a_i^+ a_j^+$ connects the ground state
to a state in which the semi-infinite chain to the right of the wall
is in an excited state which we label $\beta$ and the semi-infinite
chain to the left of the wall is in an excited state $\alpha$.  Thus
we have
\begin{equation}
E_1^{(2)} = - S^{-1}  \sum_{ijkl} \sum_{\alpha \beta}  X_{ij}^{(0)}
X_{kl}^{(0)} { \phi_\alpha^{(-\infty,0)} (i)
\phi_\alpha^{(-\infty,0)} (k) \phi_\beta^{(1,\infty)}(j)
\phi_\beta^{(1,\infty)}(l) \over \epsilon_\alpha^{(\infty)}
+ \epsilon_\beta^{(\infty)}  } \ .
\end{equation}
This process is illustrated in Fig. \ref{fig:e12}.

\begin{figure}
\centerline{\psfig{figure=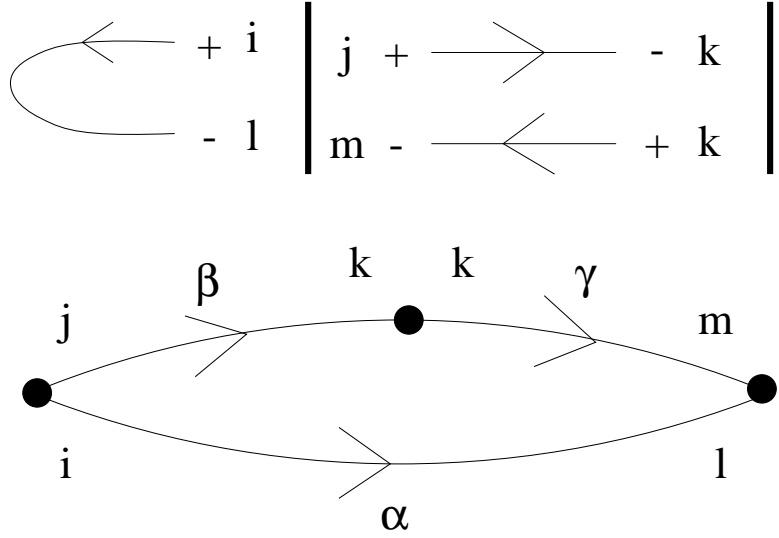,width=4in}}
\vspace{0.5cm}
\caption{Contribution to the energy of a two-wall configuration
from third-order perturbation theory.
In the real space representation excitations are created at sites
$i$ and $j$ and ultimately destroyed at sites $l$ and $m$ all
near the left wall.}
\label{fig:2w3pt}
\end{figure}

We now construct the energy of a system with only two walls,
one between sites $0$ and $1$, the other between sites $n$ and
$n+1$.  The contribution to the total
energy of this configuration from second-order perturbation
theory, denoted $E_2^{(2)}(n)$ comes from an expression
similar to Eq. (\ref{WALL1}) but which here
involves one semi-infinite chain and one block of length $n$,
\begin{equation}
E_2^{(2)} (n) = - 2  S^{-1}  \sum_{ijkl} \sum_{\alpha \beta}  X_{ij}^{(0)}
X_{kl}^{(0)} { \phi_\alpha^{(-\infty,0)} (i)
\phi_\alpha^{(-\infty,0)} (k) \phi_\beta^{(1,n)}(j)
\phi_\beta^{(1,n)}(l) \over \epsilon_\alpha^{(\infty)}
+ \epsilon_\beta^{(n)}  } \ .
\end{equation}
Here and below we include a factor of 2 because the process could
be initiated at either of the two walls.
Note that as $n \rightarrow \infty$, $E_2^{(2)} \rightarrow 2 E_1^{(2)}$,
as expected.  We will also need the contribution to the energy of
this configuration from third-order perturbation theory.  The only
process at this order is shown in Fig. \ref{fig:2w3pt} and it 
gives a contribution

\begin{equation}
\label{PERT}
E_2^{(3)}(n) = 2 \langle 0 | Y^{(0)} { Q_0 \over {\cal E}}
W^{(n)} { Q_0 \over {\cal E}} X^{(0)} | 0 \rangle =
2 S^{-3} \sum_{ijklm} X_{lm}^{(0)} W_k^{(n)} X_{ij}^{(0)}
\langle 0 | a_l a_m { Q_0 \over {\cal E}} a_k^+ a_k
{ Q_0 \over {\cal E}} a_i^+ a_j^+ | 0 \rangle \ .
\end{equation}
In Eq. (\ref{PERT}) we see that $X^{(0)}$ creates one excited
eigenstate in the semi-infinite chain to the left of the walls and also
an excited eigenstate in the down-spin block of length $n$.  That type
of reasoning allows us to rewrite Eq. (\ref{PERT}) as
\begin{equation}
E_2^{(3)} (n) = 2  S^{-1}  \sum_{ijklm} \sum_{\alpha \beta \gamma}
X_{ij}^{(0)} W_k^{(n)} X_{lm}^{(0)}
{ \phi_\alpha^{(-\infty,0)} (i) \phi_\beta^{(1,n)}(j)
\phi_\beta^{(1,n)} (k) \phi_\gamma^{(1,n)}(k)
\phi_\alpha^{(-\infty,0)} (l) \phi_\gamma^{(1,n)}(m) \over
\Bigl( \epsilon_\alpha^{(\infty)} + \epsilon_\beta^{(n)} \Bigr)
\Bigl( \epsilon_\alpha^{(\infty)} + \epsilon_\gamma^{(n)} \Bigr) } \ .
\end{equation}
Then, up to third-order perturbation contributions,
the wall potential $V_2(n)$ we wish to obtain is given by
\begin{equation}
V_2 (n) =  E_2^{(2)}(n) - 2 E_1^{(2)} + E_2^{(3)}(n) \ .
\label{eqn:v_2_n}
\end{equation}

To interpret these expressions it is convenient to express them in
terms of the Green's function, defined by
\begin{equation}
G_{ij}^{(1,n)}(E) = \sum_\alpha {\phi_\alpha^{(1,n)} (i)
\phi_\alpha^{(1,n)} (j) \over E + \epsilon_\alpha^{(n)}} \ .
\end{equation}
Thus 
\begin{equation}
E_1^{(2)} = - S^{-1} \sum_{ijkl} \sum_\alpha  X_{ij}^{(0)}
X_{kl}^{(0)} \phi_\alpha^{(-\infty,0)} (i)
\phi_\alpha^{(-\infty,0)} (k) G_{jl}^{(1,\infty)}
(\epsilon_\alpha^{(\infty)}),
\label{eqn:exx1}
\end{equation}
\begin{equation}
E_2^{(2)} (n) = - 2 S^{-1} \sum_{ijkl} \sum_\alpha  X_{ij}^{(0)}
X_{kl}^{(0)} \phi_\alpha^{(-\infty,0)} (i)
\phi_\alpha^{(-\infty,0)} (k) G_{jl}^{(1,n)}
(\epsilon_\alpha^{(\infty)}),
\end{equation}
and
\begin{equation}
E_2^{(3)} (n) = 2 S^{-3} \sum_{ijklm} \sum_\alpha 
X_{ij}^{(0)} W_k^{(n)} X_{lm}^{(0)}
\phi_\alpha^{(-\infty,0)} (i) \phi_\alpha^{(-\infty,0)} (l)
G_{jk}^{(1,n)} ( \epsilon_\alpha^{(\infty)})
G_{km}^{(1,n)} ( \epsilon_\alpha^{(\infty)}) \ .
\label{eqn:exx2}
\end{equation}

To obtain $V_2(n)$ we will have to determine
$\delta G \equiv G^{(1,n)}-G^{(1,\infty)}$.
To evaluate this quantity we need to identify the
perturbation which, when added to the unperturbed
Hamiltonian describing two independent blocks of spins, $(1,n)$
and $(n +1 , \infty)$, gives the unperturbed Hamiltonian
${\cal H}_0^{(1,\infty)}$.  This perturbation $V$ can be written
as
\begin{equation}
\label{VV}
V = -W^{(n)}+ Z^{(n)} \ ,
\end{equation}
where $Z^{(n)}$ describes hopping across the wall which is needed
to make the semi-infinite chain from a finite block of parallel spins.

We now use some results of standard perturbation theory for a
Green's function, as given, for instance, in Ref. \onlinecite{RMP}.
For this expansion
we work to lowest order in the wall perturbation, $V$ of Eq. (\ref{VV}).
We choose ${\cal H}_0$ to be the Hamiltonian for a block of $n$
parallel spins and treat $V$ perturbatively.  In first-order
perturbation theory for $V$, it is not necessary to keep $Z^{(n)}$
(and consequently ${\cal H}_0^{(n+1,\infty)}$)
because it moves an excitation to the right of the right wall
which cannot be hopped back to the $(1,n)$ block without going to higher-order
perturbation theory.  So
correct to first order in perturbation theory we have 

\begin{eqnarray}
G^{(1,\infty)}_{ij} & = & \Biggl[ \Bigl( 
E + S {\cal H}_0^{(1,n)} -S W^{(n)}
\Bigr)^{-1} \Biggr]_{ij} \nonumber \\ &=&
\Biggl[ \Bigl( E + S {\cal H}_0^{(1,n)} \Bigr)^{-1} \Biggr]_{ij}
+ \sum_k \Biggl[ \Bigl( E + S {\cal H}_0^{(1,n)} \Bigr)^{-1}
\Biggr]_{ik} W^{(n)}_k
\Biggl[  \Bigl( E + S {\cal H}_0^{(1,n)} \Bigr)^{-1}
\Biggr]_{kj} \nonumber
\\ &=&
G^{(1,n)}_{ij}+  \sum_k  G^{(1,n)}_{ik}W^{(n)}_k G^{(1,n)}_{kj} \ ,
\label{eqn:gr}
\end{eqnarray}
\noindent where $W_k^{(n)}$ is defined in Eq. (\ref{eqn:ws}).
Thus  using Eqs. (\ref{eqn:v_2_n}), (\ref{eqn:exx1})-(\ref{eqn:exx2}) and
(\ref{eqn:gr})  we have  the result
\begin{equation}
\label{V2NEQ}
V_2(n) = 4S^{-1}  \sum_{ijklm} \sum_\alpha X_{ij}^{(0)} W_k^{(n)}
X_{lm}^{(0)} \phi_\alpha^{(-\infty,0)}(i) \phi_\alpha^{(-\infty,0)}(l)
G_{jk}^{(1,n)} ( \epsilon_\alpha^{(\infty)})
G_{km}^{(1,n)} ( \epsilon_\alpha^{(\infty)}) \ .
\label{eqn:v2xwx}
\end{equation}
Evaluating this when $J_1=2J_2$ we obtain [writing $G$ for
$G^{(1,n)}(\epsilon_\alpha^\infty )$ and $\phi$ for $\phi^{(-\infty,0)}$]
\begin{eqnarray}
V_2(n)&=&{4J_2^3 \over S} \sum_\alpha \Biggl\{ \phi_\alpha (0)^2
\Biggl[ \Biggl( G_{2,n-1} - 2G_{1,n-1}\Biggr)^2 
-\Biggl( G_{2,n} - 2G_{1,n}\Biggr)^2 \Biggr] \nonumber \\ &&
\ + 2\phi_\alpha(-1)\phi_\alpha(0) \Biggl[
\Biggl( G_{2,n-1}-2G_{1,n-1} \Biggr) G_{1,n-1}
- \Biggl( G_{2,n} - 2G_{1,n}\Biggr) G_{1,n} \Biggr] \nonumber \\ && \ 
+ \phi_\alpha(-1)^2 \Biggl( G_{1,n-1}^2 - G_{1,n}^2 \Biggr) \Biggr\} 
\\ &=& \label{V2GEN}
{4J_2^3 \over S} \sum_\alpha \Biggl\{ \Biggl[ \phi_\alpha(0)
\Biggl( G_{2,n-1} -2 G_{1,n-1} \Biggr) + \phi_\alpha(-1) G_{1,n-1}
\Biggr]^2 \nonumber \\ && \ - \Biggl[ \phi_\alpha(0) \Biggl( G_{2,n}
-2 G_{1,n} \Biggr) + \phi_\alpha(-1) G_{1,n} \Biggr]^2 \Biggr\} \ .
\label{eqn:Vngreen}
\end{eqnarray}

We will evaluate this with successively increasingly accurate
approximations for large $n$.  For small $n$ it is certainly
correct to replace $\epsilon_\alpha^{(\infty)}$ by
$2D'\equiv 2\tilde D + 2J_1-2J_2= 2 \tilde D + 2 J_2$, since
corrections will be proportional to $J/D'$ with a bounded coefficient.
For the moment we continue to use this approximation even for
large $n$. With this approximation, the sum over
$\alpha$ in Eq. (\ref{eqn:v2xwx}) yields
\begin{equation}
\label{COMPLETE}
\sum_\alpha \phi_\alpha^{(-\infty,0)}(i) \phi_\alpha^{(-\infty,0)}(l)
= \delta_{i,l} \ .
\end{equation}
 We refer to this as the nonpropagation approximation,
since it  amounts to setting the off-diagonal elements of ${\cal
H}_0^{(-\infty,0)}$ to zero, forcing $i$ and $l$ to coincide.

Within this approximation and writing writing
$G$ for $G^{(1,n)}(2D')$, we find that
\begin{equation}
\label{V2EQ}
V_2(n)  = V_A + V_B 
\end{equation}
where
\begin{equation}
\label{VAEQ}
V_A =  {4J_2^3 \over S} \biggl( G_{2,n-1} - 2G_{1,n-1} \biggr)^2 \ , 
\end{equation}
\begin{equation}
\label{VBEQ}
V_B =  {4J_2^3 \over S} \biggl( 4 G_{1,n} G_{1,n-1} - 5 G_{1,n}^2
\Biggr) \ .
\end{equation}

In Appendix B we give an essentially exact evaluation of the
required $G$'s, apart from an overall scale factor which we only
obtain approximately.  This evaluation leads to the result
\begin{eqnarray}
\label{V2RES}
V_2 (n) &=& { 16 D' \over S \lambda^n }
\Biggl( \sin^2 [ n \delta + 4\delta^3 ] -  (3/8) (J_2/D')^3 \Biggr)
\ , \ \ \ \ n \ {\rm even} \ ; \nonumber \\
&=& { 16 D' \over S \lambda^n }
\Biggl( \cos^2 [ n \delta + 4\delta^3 ] -  (3/8) (J_2/D')^3 \Biggr)
\ , \ \ \ \ n \ {\rm odd} \ ;
\end{eqnarray}
where, to leading order in $J_2/(4D')$,
$\lambda^{-1} = \delta^2 = J_2/(4D')$.  Here $V_A$ gives rise to
the term involving the square of the trigonometric function and,
when $V_A=0$,
\begin{equation}
V_B = - { 6 D' \over S \lambda^n } \left( {J_2 \over D' } \right)^3 \ .
\end{equation}
For small $n$ these expressions reduce to our previous results 
(\ref{V2odd}) and (\ref{V2even})  at
leading order in $J_2/D'$, in which case $V_B$ is irrelevant.

We now discuss the interpretation of these results.  For the moment
let us ignore completely the term $V_B$.
When $V_2(n)$ is nonmonotonic, as we found here, an elegant
graphical construction which yields the phase diagram was
suggested by Szpilka and Fisher [\onlinecite{MEFXS}].   This
proceeds by drawing  the lower convex envelope of the points
$V_2(n)$ versus $n$.  Points on the convex envelope are the
allowed stable phases (assuming no further bifurcation due
to $V_3$).  For this construction it is important to distinguish
the case when $V_2(n)$ becomes negative. If  this occurs,  
then there will be a first-order transition from $n_0$ to
$n=\infty$ where $n_0$ is the value of $n$ for which $V_2(n)$
attains its most negative value.
On the other hand, if $V_2(n)$ is positive for all $n$, then
one has an infinite devil's staircase, with no bound on the
allowed values of $n$.  Accordingly, it is obviously important to
ascertain whether or not $V_2(n)$ is positive definite.
Eq. (\ref{V2RES}) suggests that $V_2(n)$ can become
negative when  $(n \delta + 4 \delta^3 )/(2\pi)$ is sufficiently
close to an integer. However the approximations inherent in its
derivation may alter this conclusion.

\subsection{Effect of Allowing Propagation of the Left Excitation}

To determine whether or not an unending devil's staircase  actually
exists in the phase diagram, it is necessary to assess the validity
of the nonpropagation approximation.  We now avoid the approximate
treatment of Eq. (\ref{V2NEQ}) in which we replaced
$\epsilon_\alpha^{(\infty)}$ by $2D'$.
We write Eq. (\ref{V2NEQ}) as
\begin{equation}
V_2(n) = 4S^{-1}  \sum_{ijklm} \sum_\alpha X_{ij}^{(0)} W_k^{(n)}
X_{lm}^{(0)} \phi_\beta^{(1,n)}(j) \phi_\beta^{(1,n)}(k)
\phi_\gamma^{(1,n)}(k) \phi_\gamma^{(1,n)}(l) Y \ ,
\end{equation}
where
\begin{eqnarray}
\label{YY}
Y & \equiv & \sum_\alpha { \phi_\alpha^{(-\infty,0)} (i)
\phi_\alpha^{(-\infty,0)} (l)
\over [\epsilon_\alpha^{(\infty)} + \epsilon_\beta ^{(n)}]
[\epsilon_\alpha^{(\infty)} + \epsilon_\gamma ^{(n)}]} \nonumber \\
&=& \sum_\alpha { \phi_\alpha^{(-\infty,0)} (i)
\phi_\alpha^{(-\infty,0)} (l) \over [2D' + \epsilon_\beta ^{(n)}]
[2D' + \epsilon_\gamma ^{(n)}]} \Biggl( 1 -
{\epsilon_\alpha^{(-\infty,0 )} -2D' \over 2D' +
\epsilon_\beta^{(n)} } - {\epsilon_\alpha^{(-\infty,0 )} -2D' 
\over 2D' + \epsilon_\gamma^{(n)} } \dots \Biggr) \nonumber \\
&\equiv& Y_0 + \sum_\alpha \phi_\alpha^{(-\infty,0)} (i)
\phi_\alpha^{(-\infty,0)} (l)
\Biggl( \epsilon_\alpha^{(-\infty,0 )} - 2D' \Biggr)
\Biggl[ { 1 \over 2} { \partial \over \partial D' } \Biggl(
{1 \over 2 D' + \epsilon_\beta^{(n)} } {1 \over 2D' +
\epsilon_\gamma^{(n)} } \Biggr) \Biggr] \nonumber \\
& \equiv & Y_0 + \delta Y \ .
\end{eqnarray}
Keeping only the term $Y_0$ leads to the nonpropagation approximation,
and thence to Eq. (\ref{COMPLETE}) and the results of Eq. (\ref{V2RES}).

\begin{figure}
\centerline{\psfig{figure=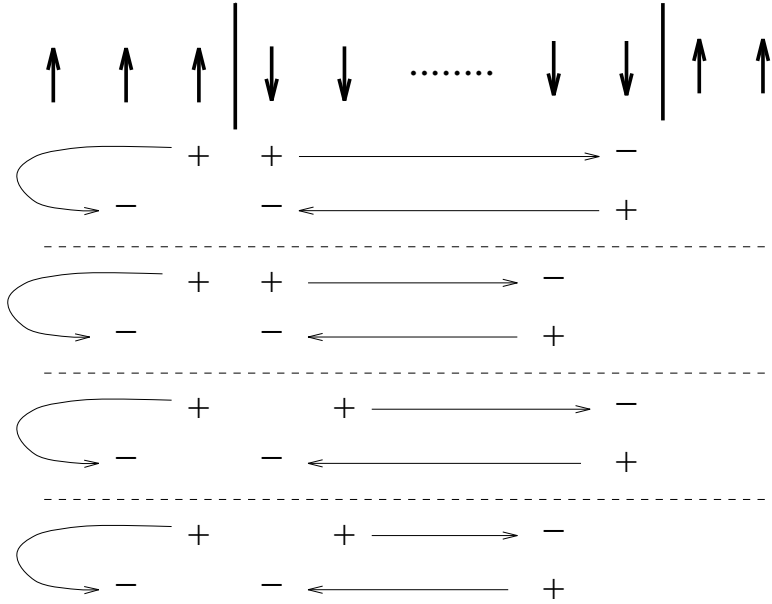,width=4in}}
\vspace{0.5cm}
\caption{Contributions to $\delta V$ as per Eq.  (\protect\ref{DV})
in which the excitation to the left of the wall propagates.
Nonpropagating contributions are similar to those in
Fig. \protect\ref{fig:2wnoe}.  Note 
that only the term in which $i$ and $l$ are nearest neighbors
actually appears.  Each diagram occurs four times: twice at one
wall by interchanging creation and annihilation sites, and twice
for interchanging the roles of the two walls.}
\label{fig:dv}
\end{figure}

We now analyze the effect of $\delta Y$.  For that purpose
we use the fact that the eigenfunctions
satisfy Eq. (\ref{EIGEQ}).  For sites $i$ near the wall,
(i.e., $i=0$ and $i=-1$) Eq. (\ref{EIGEQ}) yields [omitting the cumbersome
superscripts $(-\infty,0)$]
\begin{equation}
\Bigl( \epsilon_\alpha -2 D' \Bigr) \phi_\alpha(0) =
(J_2-J_1) \phi_\alpha (0) - J_1 \phi_\alpha (-1) + J_2 \phi_\alpha (-2)
\end{equation}
\begin{equation}
\Bigl( \epsilon_\alpha -2 D' \Bigr) \phi_\alpha(-1) =
J_2 \phi_\alpha (-1) - J_1 \phi_\alpha (0) -J_1 \phi\alpha(-2)
+ J_2 \phi_\alpha (-3) \ .
\end{equation}
Using these equations and also Eq. (\ref{COMPLETE}), we get
\begin{eqnarray}
\delta Y = & \Biggl[ & (J_2-J_1) \delta_{i,0} \delta_{l,0}
+ J_2 \delta_{i,-1} \delta_{l,-1} - J_1 \delta_{i,0} \delta_{l,-1}
- J_1 \delta_{i,-1} \delta_{l,0} \Biggr] \nonumber \\
&\times & \Biggl[ { 1 \over 2} { \partial \over \partial D' } \Biggl(
{1 \over 2 D' + \epsilon_\beta^{(n)} } {1 \over 2D' +
\epsilon_\gamma^{(n)} } \Biggr) \Biggr] \ .
\end{eqnarray}
$\delta Y$ leads to contributions with $i\not= l$ shown in Fig. \ref{fig:dv},
and also with $i=l$ which are not shown but are similar to those of
Fig. \ref{fig:2wnoe}.  If $G$ denotes $G^{(1,n)}(2D')$, then we have,
from Eq. (\ref{eqn:v2xwx}) a correction to the two-wall interaction of 
\begin{eqnarray}
\label{DV}
\delta V_2(n) & = & {2 S^2} { \partial \over \partial D'} \sum_{ijklm}
X_{i,j}^{(0)} W_k^{(n)} X_{l,m}^{(0)} G_{j,k} G_{k,m} \nonumber \\
&& \ \ \ \times \Biggl[ (J_2-J_1) \delta_{i,0} \delta_{l,0}
+ J_2 \delta_{i,-1} \delta_{l,-1} - J_1 \delta_{i,0} \delta_{l,-1}
- J_1 \delta_{i,-1} \delta_{l,0} \Biggr] \nonumber \\ 
&=& {2 \over S} {\partial \over \partial D'} \Biggl\{
4J_1^2J_2 \Bigl( J_2 G_{1,n-1}^2 + (J_2-J_1) G_{1,n}^2 \Bigr)
\nonumber \\ && \ \
- 4J_1 J_2^2 \Bigl( J_2 G_{2,n-1}G_{1,n-1} + (J_2-J_1)G_{1,n}
G_{2,n} \Bigr)
+ 2J_2^3 \Biggl( J_2 G_{1,n-1}^2 + (J_2-J_1) G_{1,n}^2 \Biggr)
\nonumber \\ && \ \ 
+ 2 (J_2-J_1)J_2^2 \Biggl( J_2 G_{2,n-1}^2 + (J_2-J_1)G_{2,n}^2 \Biggr) +
2 (J_2-J_1)J_1^2 \Biggl( J_2 G_{1,n-1}^2 + (J_2-J_1)G_{1,n}^2 \Biggr)
\nonumber \\ && \ \ -
4(J_2-J_1)J_1J_2 \Biggl( J_2 G_{1,n-1} G_{2,n-1}
+ (J_2-J_1) G_{1,n} G_{2,n} \Biggr) \Biggr\} \ .
\end{eqnarray}

We simplify this by setting $J_1=2J_2$.  Then
\begin{equation}
\delta V_2(n) = {2J_2^4 \over S} {\partial \over \partial D'}
\Biggl( -2 G_{2,n-1}^2 + 12 G_{1,n-1}^2 - 10G_{1,n}^2 \Biggr) \ . 
\end{equation}
The dominant contribution comes from the first term.  To evaluate
this expression, it is necessary to develop an expression for
$G_{2,n-1}$.  Using Eq. (\ref{GEQS}) of Appendix B, we write
\begin{equation}
\Delta_n (-1)^{n+1} G_{2,n-1} = C^2 d_{n-3}
= J_2^{n-1} (C/J_2)^2 y^{n-3} Q_{n-3} = J_2^{n-1} y^{n+1} Q_{n-3} \ ,
\end{equation}
where $y = \sqrt {4D'/J_2}$ and we
set $C=4D'$.  The calculations for $n$ odd and
$n$ even are similar.  Here we do them only for $n$ even, in which case
\begin{equation}
G_{2,n-1}^2 = { J_2^{2n-2} \over \Delta_n^2 } y^{2n+2}
\sin^2 (n \delta - 2 \delta ) \ .
\end{equation}
To take the derivative note that $G_{2,n-1}^2 \sim D'^{-n-1} \sin^2
(n \delta - 2 \delta)$ and $\delta \sim D'^{-1/2}$.  Thus we have
\begin{equation}
{d G_{2,n-1}^2 \over dD' } = { J_2^{2n-2} \over \Delta_n^2 }
y^{2n+2} \Biggl( -{(n+1) \over D'} \sin^2 (n \delta - 2 \delta)
- 2 \sin (n \delta - 2 \delta ) \cos (n \delta - 2 \delta) 
{(n-2) \delta \over 2 D'} \Biggr) \ .
\end{equation}
Then we obtain, for $n \gg 1$ and $n \delta/\pi$ an integer,
\begin{equation}
{d G_{2,n-1}^2 \over dD'} = - 2 {J_2^{2n-2} y^{2n+2} \over \Delta_n^2 } 
{n \delta^2 \over D^\prime} \ ,
\end{equation}
so that
\begin{equation}
\delta V_2(n) = {8J_2^2 y^2 \over D'S}  n \delta^2 {J_2^{2n} y^{2n} \over
\Delta_n^2} 
= {8J_2^2 y^2 \over D'S} n \delta^2 \left( {y \over \lambda} \right)^{2n}
\approx {8J_2^2 y n \delta \over D'S \lambda^n } 
\ .
\end{equation} 
Since $n \delta \geq 1$ when $V_A=0$, this term is larger than
$V_B$ by at least
$(D'y/J_2) \sim (4D'/J_2)^{3/2}$. 
Since this term is positive, we see that allowing for the left-hand
excitation to propagate
leads to a correction to $V_A$ which is much more important than $V_B$.
In other words, this more accurate evaluation gives
$V_2(n)=V_A + \delta V_2(n)$.
This result relies on the validity of the expansion in Eq. (\ref{YY}),
the precise condition for which is not obvious.  However, when $n$
gets sufficiently large, this expansion breaks down and the
considerations in the next subsection become necessary.

\subsection{Large $n$ Limit with Propagation}

The expansion that we have used in Eq. (\ref{YY}) implicitly
assumes that the Green's function has a weak dependence on
energy.  That is true as long as $n$ is small enough.  But
when $n$ becomes arbitrarily large, then there must exist
a regime in which the right-hand side of Eq. (\ref{V2GEN})
is dominated by the largest term in the sum over $\alpha$.
If it were correct to keep only a single value of $\alpha$,
then it would be possible to fix $D'$, so that the first
square bracket in Eq. (\ref{V2GEN}) would vanish and $V_2(n)$
would be negative.  This reasoning is not correct, however as the
analysis in Appendix C shows. 
Even for large $n$ the sum over $\alpha$ has a width in $\alpha$
of order $\sqrt{1/n}$ which prevents
$(G_{2,n-1}-2G_{2,n})^2$ from being fixed to be precisely
zero.  We therefore
conclude that for the one-dimensional system of walls, $V_2(n)$
does remain positive for $n \rightarrow \infty$.  It seems
unlikely that in a crossover between the regimes we have considered
$V_2(n)$ would become negative.  So we conclude that for the
one-dimensional problem $V_2(n)$ remains positive and there is
no cutoff in the Devil's staircase for the phase diagram.

\subsection{Large $n$ Limit For Three-Dimensional Systems}

In the discussion up to now, we have treated the three-dimensional
system as if it were a one-dimensional system in which planar walls
separate up-spin segments from down-spin segments.  Here we give
a brief argument which suggests that these one-dimensional
results continue to hold for the three-dimensional system.
One way to phrase the argument is to note that when $D'$ is
large compared to the $J$'s, we are far from criticality.
The correlation length (of order $| \ln (J/D^\prime)|^{-1}$) is
very short.  Thus, entropic effects of longer paths are
strongly cut off by the correlation length.  Here we indicate
the nature of a formal argument of this type. 

\begin{figure}
\centerline{\psfig{figure=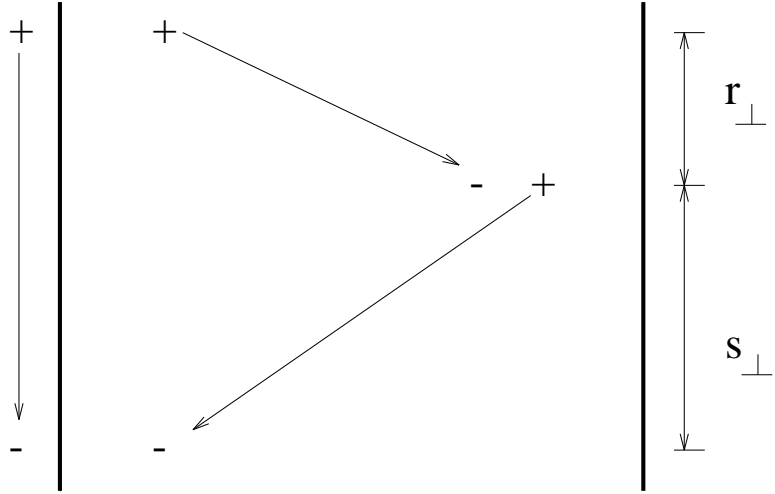,width=4in}}
\vspace{0.5cm}
\caption{Leading contribution to $V_2(n)$ for a three-dimensional
system.  Here $r_\perp$ and $s_\perp$ denote vectors in the plane
of the wall.}
\label{fig:3d}
\end{figure}

To analyze the three-dimensional case, we consider only the
dominant term, illustrated in Fig. \ref{fig:3d}.  It gives rise to the
contribution
\begin{equation}
\delta V_2(n) = {4J_2^3 \over S} \sum_{r_\perp, s_\perp}
\sum_{\alpha, q_\perp} \phi_{\alpha, q_\perp} (0;0)
\phi_{\alpha, q_\perp} (0;r_\perp+s_\perp)
G_{2,n-1}(r_\perp ; \epsilon_{\alpha, q_\perp}) G_{2,n-1}(s_\perp ;
\epsilon_{\alpha,q_\perp }) \ ,
\end{equation}
where we omit the superscripts.  The subscripts on $\phi$
are the quantum number, $\alpha$, associated with the coordinate
perpendicular to the wall and the wavevector $q_\perp$ associated
with the transverse coordinates.  The arguments of $\phi$ are the
coordinate perpendicular to the wall and the vector displacement
in the plane of the wall.  The arguments of $G$ are the displacement
in the plane of the wall and the energy.  Considering only the
dependence of $\phi$ and $\epsilon$ on wavevector we obtain
\begin{equation}
\delta V_2(n) \sim \sum_{r_\perp, s_\perp , q_\perp}
\exp [ i {\bf q_\perp} \cdot {(\bf r_\perp + s_\perp)} ]
G_{2,n-1}(r_\perp;\epsilon_{q_\perp}) G_{2,n-1}(s_\perp;
\epsilon_{q_\perp})  \ .
\end{equation}
In terms of Fourier transformed variables for coordinates in the
plane of the wall (indicated by overbars), we have
\begin{equation}
\label{V23D}
\delta V_2(n) \sim \sum_{q_\perp}
\Bigl[ \bar G_{2,n-1}(q_\perp;\epsilon_{q_\perp})\Bigr]^2 \ .
\end{equation}
But this is again the type of expression analyzed in Appendix C.
So we conclude that for the three-dimensional system $V_2(n)$
is also always positive.

\section{Discussion}

The aim in this paper has been to demonstrate how quantum fluctuations
can lead to interactions between domain walls and hence stabilize
long-period phases in the vicinity of a multiphase point where the
intrinsic wall energy is small.

We first considered a Heisenberg model with strong uniaxial spin
anisotropy $D$ and an interface pinned to a surface by a bulk magnetic
field $h$. A perturbation expansion in $D^{-1}$ was used to show that
the wall-interface interaction is repulsive and hence that the
interface unbinds from the surface through an infinite number of
layering transitions as $h$ passes through 0.

The bulk of the paper was devoted to describing the behavior of the
Heisenberg model with first- and second-neighbor competing
interactions and uniaxial anisotropy $D$ near the ANNNI model limit
$D=\infty$. This model has a multiphase point for sufficiently large
$D$ which is split by quantum fluctuations to give a sequence of
long-period commensurate phases $\langle 2 \rangle$, $\langle 3
\rangle$, $\langle 4 \rangle$, ... $\langle n \rangle$ ... .

The phase sequence could be established for $n$ not too large by a
calculation of two-wall and three-wall interactions using perturbation
theory with $D^{-1}$ as a small parameter. A discussion of correction
terms important for large $n$ was given from which we concluded that,
unlike for the ANNNI model, the sequence of phases is infinite.
The reason this model is different in this regard from the 
ANNNI model is an inherently quantum one: for one wall to
indirectly interact with another an excitation has to propagate
from one wall to the other AND return.  Thus the interaction
in the quantum case is proportional to the square of a
oscillatory Green's function whereas in the ANNNI model the analogous
function appears linearly.  As a consequence of this oscillation,
the phases come in the sequence $n \rightarrow n+1$ or
$n \rightarrow n+2$, depending on the value of $n$. 
In the latter case, we did not explicitly investigate the
stability of the phase diagram, but a cursory analysis leads
us to believe that the function $F(n,n+2)$
analogous that in Eq. (\ref{FEQ}) is negative.

Similar behavior is observed in both the interface model
[\onlinecite{DuxY}] and the ANNNI model
[\onlinecite{MEFXS},\onlinecite{MEFWS}] for finite temperatures.
Here thermal fluctuations replace quantum fluctuations in mediating
the domain wall interactions.
Although long-period phases are stabilized in the ANNNI case, the
qualitative form of the phase sequence is very different to that
discussed in this paper.  The stable phases are $\langle 2^k 3 \rangle$,
$k=0,1,2,\ ...\ k_{\rm max}$ with $ k_{\rm max} \to \infty$ as
$T \to 0$. Mixed phases $\langle 2^k 3 2^{k-1} 3 \rangle$
also appear [\onlinecite{MEFXS}].

A third mechanism that can split the degeneracy at both a bulk
[\onlinecite{SY}] and an interface [~\onlinecite{MY}~] multiphase point
is the softening of the spins themselves; a non-infinite spin
anisotropy. This does not occur for the ANNNI model where there is a
finite energy barrier for the spins to move from their positions at
$D=\infty$. However, for a similar model with 6-fold anisotropy an
infinite number of phases become stable near the multiphase point as
$D$ is reduced from infinity. [\onlinecite{SY}]

\acknowledgments

JMY is supported by an EPSRC Advanced Fellowship and CM by an
EPSRC Studentship and the Fondazione "A. della Riccia", Firenze.
ABH acknowledges the receipt of an EPSRC Visiting Fellowship.
Work done at Tel Aviv was also supported in part by the USIEF
and the US-Israel BSF.

\vskip 2.2 cm

\appendix

\section{Stability of the $\langle 2 \rangle$:$\langle 3 \rangle$
boundary}

In general $F(2n,2n+1)$ is ${\cal O} (1/\tilde{D}^{4n-1})$. However, for $n=1$
the term ${\cal O} (1/\tilde{D}^{3})$ is accidentally zero and terms ${\cal O}
(1/\tilde{D}^4)$ must be retained. This leads to a lengthy calculation. We
now calculate $F(2,3)$ explicitly, considering each order of
perturbation theory in turn.

\underline{Second-order perturbation theory}

Contributions arise from diagrams spanning a wall which are created
and then immediately destroyed (as in the example of Fig. \ref{fig:2pt}). The
contribution to $V_3(2,2)$ comes from both first and second neighbor
excitations. That to $V_3(2,3)$ is just from second neighbors because
the energy denominator of the first neighbor excitation does not
depend on both $\sigma$ and $\eta$.
For the same reason there is no contribution at all to
$V_3(3,3)$.

Using a subscript $2$ to indicate that we are considering only the
terms arising from second order perturbation theory one obtains
${\cal O}(1/\tilde{D}^4)$
\begin{eqnarray}
V_{2}(2,2)=&& {8J_2^2 \over (4\tilde{D})^3S} [ -J_1^2 + 2 J_2(J_1-J_2)]
+ {48J_2^3 \over (4\tilde{D})^4S} [ - J_1 J_2+2J_2^2],\label{s1}\\
V_{2}(2,3)=&&-{8J_2^4 \over (4\tilde{D})^3S} +
{48J_2^4(J_2+J_1) \over (4\tilde{D})^4S},\label{s2} \\
V_{2}(3,3)=&&0.
\end{eqnarray}

\underline{Third-order perturbation theory}

\begin{figure}
\centerline{\psfig{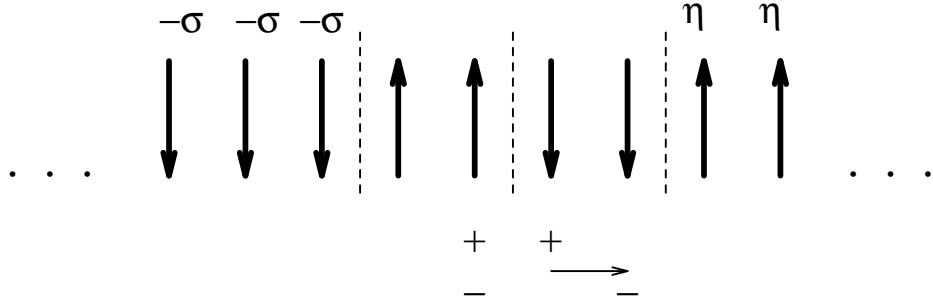}}
\vskip 0.5cm
\caption{Example of a term contributing to $V_3(2,2)$ in third order
perturbation theory.}
\label{fig:v322}
\end{figure}

Contributions to $V_3(2,2)$ in third order perturbation theory arise
from diagrams like that shown in Fig. \ref{fig:v322}. Recalling that the spins on
either side of the wall can hop, and that the initial excitation can
be between second neighbors, with a subsequent hop to first
neighbors gives
\begin{equation}
V_3(2,2)={48 J_1^2J_2^2 (2J_2-J_1) \over (4\tilde{D})^4S }.
\label{s3}
\end{equation}
Similar diagrams contribute to $V_3(2,3)$. The hop must lie within the
domain of 3 spins
\begin{equation}
V_3(2,3)={24 J_1^2J_2^3  \over (4\tilde{D})^4S }.
\label{s4}
\end{equation}
There is no contribution $V_3(3,3)$.\\

\underline{Fourth-order perturbation theory}

We first consider processes which are proposrtional to  $V_{||}^{2}
V_{\not{\parallel}}^{2}$.
As we discussed in the text, to lowest order in $J_2/D$ we do
not need to consider processes which hop beyond the wall.
However, since the calculation of $F(2,3)$ requires a
calculation of $V_3(2,2)$ and $V_3(2,3)$ including the first
higher-order corrections, we need to keep such processes.

\begin{figure}
\centerline{\psfig{figure=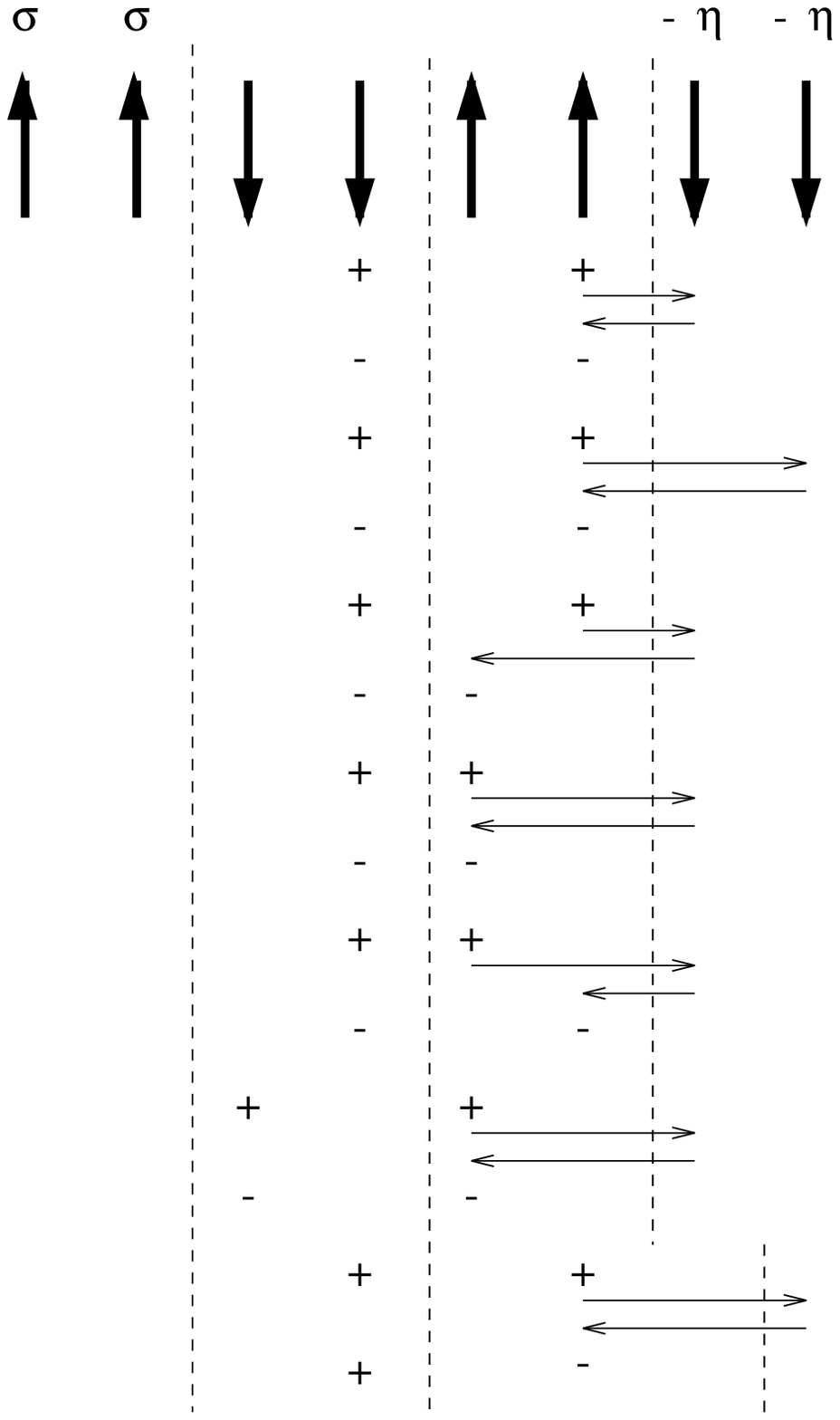,height=6.0in}}
\vskip 0.5cm
\caption{Processes which can not occur when the wall
is as shown and which therefore carry a factor
$\delta_{\eta,-1}$
(when the
wall is absent $\eta=-1$ and these processes are allowed).  The first
six diagrams contribute to $V_3(2,2)$ and the last one to $V_3(2,3)$.
In the last diagram the right-hand block contains three down spins.}
\label{fig:y}
\end{figure}

We now evaluate contributions from such processes, which we show in
Fig. \ref{fig:y}.  First of all, since these processes only exist in the
{\it absence} of the right-hand wall, they all carry a factor
$\delta_{\eta,-1}$.  Secondly their overall sign is negative for
even-order (fourth-order) perturbation theory.   Also, the contributions
to $V_3(2,2)$ carries a factor of 2 to account for the mirror
image diagrams.

Thus from the first diagram of Fig. \ref{fig:y}, we get $E_k(\sigma ,\eta)$ of
Eq. (\ref{CONN}) as
\begin{equation}
E_3(\sigma , \eta) = - 2 \delta_{\eta , -1} J_1^2 J_2^2
({\cal E}_1 {\cal E}_2 {\cal E}_3)^{-1} \ ,
\end{equation}
where ${\cal E}_i$ is an energy denominator.  We have
that ${\cal E}_i \sim 4 \tilde D + {\rm O} (J)$.  In
particular, we need to include the dependence of ${\cal E}_i$ on
$\sigma$, which we deduce from Fig. \ref{fig:z}.

For the present purposes it suffices to set ${\cal E}_i = 4 \tilde D$
and $d {\cal E}_i /d\sigma = J_2$ for all diagrams of Fig. \ref{fig:y},
except the last one, for which $d {\cal E}_i /d\sigma = (J_2 -J_1)$.
Thus for the first diagram of Fig. \ref{fig:y} we have
\begin{equation}
\sum_{\sigma \eta} \sigma \eta E_3(\sigma , \eta) = - 12 J_1^2 J_2^2
(4 \tilde D)^{-4} ( d {\cal E} /d \sigma ) \ .
\end{equation}

Indicating with $\delta V_4(2,2)$ and $\delta V_4(2,3)$ the total
contribution to $V_4(2,2)$ and $V_4(2,3)$ from  the diagrams of
Fig. \ref{fig:y} one has

\begin{eqnarray}
\delta V_4(2,2) &=& - (J_2^3/\tilde D^4) \Bigl( 48 J_1^2 +24 J_2^2
- 12 J_1 J_2 \Bigr) \ , \label{eqn:v4u} \\
\delta V_4(2,3) &=& - (6 J_2^5/\tilde D^4) \ . \label{eqn:v4d}
\
\end{eqnarray}

\begin{figure}
\centerline{\psfig{figure=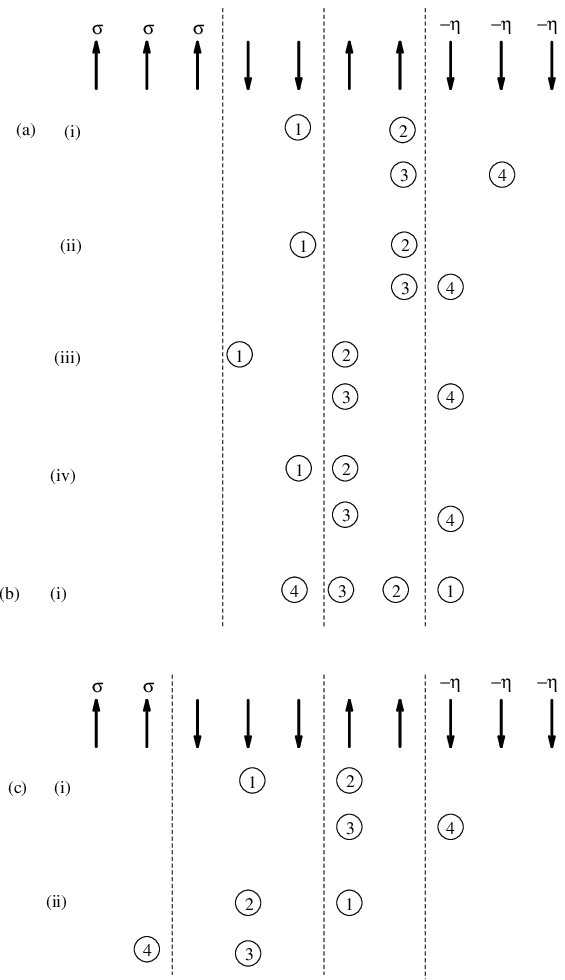,height=6.0in}}
\vskip 0.5cm
\caption{Terms which contribute to (a) $V_4^{(1)}(2,2)$, (b)
$V_4^{(2)}(2,2)$, (c) $V_4^{(3)}(2,2)$. The figures indicate which
spins are excited. The way in which all possible orderings of the
excitations are accounted for is described in the text (see equations
(A8) and (A9) for diagrams (a) and (c) and equation (A10) for diagram (b)). In
cases (a) and (b) the diagrams which are mirror images in the center
wall must also be accounted for by including a factor of 2.}
\label{fig:v4}
\end{figure}

However, one also needs to consider terms proportional to
$V_{\not{\parallel}}^{4}$ 
where two pairs of excitations are created and destroyed which do indeed turn
out to be important. Consider first a set of four spins $n_i$ at sites
$i$ and the following processes
\newcounter{count}
\begin{list}
{(\roman{count}) }{\usecounter{count} }
\item $n_1,n_2$ excited, $n_1,n_2$ destroyed,
$n_3,n_4$ excited, $n_3,n_4$ destroyed
\item $n_1,n_2$ excited, $n_3,n_4$ excited,
$n_3,n_4$ destroyed, $n_1,n_2$ destroyed
\item $n_1,n_2$ excited, $n_3,n_4$ excited,
$n_1,n_2$ destroyed, $n_3,n_4$ destroyed
\item $n_3,n_4$ excited, $n_3,n_4$ destroyed,
$n_1,n_2$ excited, $n_1,n_2$ destroyed
\item $n_3,n_4$ excited, $n_1,n_2$ excited,
$n_1,n_2$ destroyed, $n_3,n_4$ destroyed
\item $n_3,n_4$ excited, $n_1,n_2$ excited,
$n_3,n_4$ destroyed, $n_1,n_2$ destroyed

\end{list}
 We will be interested in the cases shown in Fig. \ref{fig:v4}
 where $n_1$ and $n_2$ must be first or
second neighbors straddling one wall and similarly for $n_3$ and $n_4$
with respect to the other wall.  Except for the possibility that
$n_2=n_3$, all the $n$'s are distinct.
Because we are working to linear order in $1/S$ (that is
ignoring terms higher than quadratic in the boson Hamiltonian) the
energy denominator depends on position on the lattice but not on the
position of the other excitations. Hence the energy denominators are
simply the sum of the energies of the 
excited spins relative to the ground state
energy. We denote them by $E_{ijk\ldots } $ when spins $i,j,k \ldots$
are excited. Noting that the matrix elements, say $M$, are common to
all processes (i)--(vi) we are now in a position to write down the
contribution from these diagrams to fourth order in perturbation
theory

\begin{equation}
V_4={M \over S} \left({1 \over E_{12}^2E_{34}}
                       +{1 \over E_{12}E_{34}^2}
                      -{\xi \over E_{12}E_{1234}E_{12}}
                       -{2 \xi \over E_{12}E_{1234}E_{34}}
                      -{\xi \over E_{34}E_{1234}E_{34}}
                       \right)
\end{equation}
where $\xi=2$ if two excitations are present at the same site (Bose
statistics) and $\xi=1$ for all spins distinct.

Using $E_{1234}=E_{12}+E_{34}$
\begin{equation}
V_4={M\over S}(1-\xi) {E_{12}+E_{34} \over E_{12}^2E_{34}^2}.
\end{equation}
Putting $\xi=1$ it is immediately apparent that there is no
contribution from diagrams for which all $n_i$ are different. 
There are however
terms ${\cal O}(1/\tilde{D}^4)$ when $\xi=2$. Diagrams of this type which contribute to
$V_3(2,2)$ are shown in Fig. \ref{fig:v4}a. 
Only terms with $\eta =1$ give a contribution different from $E_0$. Therefore when the sum over
$\sigma$ is taken the term proportional to $\sigma$ is the lowest
order which survives. Including a factor $2$ for diagrams symmetric
with respect to reflection in the center wall of Fig. \ref{fig:v4}a one obtains
\begin{equation}
V_4^{(a)}(2,2)={12 \over (4 {\tilde D})^4S}\biggl( 2J_2^5-J_1J_2^4+2J_1^2J_2^3
\biggr),
\label{s5}
\end{equation}
where the superscript indicates a contribution of type
a in Fig. \ref{fig:v4}.

Similarly the contributions of this type to $V_3(2,3)$ is shown
in Fig. \ref{fig:v4}c. They give
\begin{equation}
V_4^{(c)}(2,3)={12 J_2^5 \over (4 {\tilde D})^4 S}.
\label{s6}
\end{equation}

There is one further contribution to $V_4(2,2)$. Consider the
following order of excitation of four spins
\begin{list}
{(\roman{count}) }{\usecounter{count} }
\item $n_1,n_2$ excited, $n_3,n_4$ excited,
$n_2,n_4$ destroyed, $n_1,n_3$ destroyed
\item $n_1,n_2$ excited, $n_3,n_4$ excited,
$n_1,n_3$ destroyed, $n_2,n_4$ destroyed
\item $n_3,n_4$ excited, $n_1,n_2$ excited,
$n_1,n_3$ destroyed, $n_2,n_4$ destroyed
\item $n_3,n_4$ excited, $n_1,n_2$ excited,
$n_2,n_4$ destroyed, $n_1,n_3$ destroyed
\end{list}
The pairs $(n_1,n_2)$, $(n_3,n_4)$, $(n_1,n_3)$, $(n_2,n_4)$,
must all be first or second neighbors spanning a wall. This means
that the only contribution of this type is to $V_4(2,2)$
and is  shown in Fig. \ref{fig:v4}b. 
Proceeding as before the sum of all orderings gives
\begin{equation}
V_4^{(b)}=- {2 M\over S} {E_{1234} \over E_{13}E_{24}E_{12}E_{34}}
\end{equation}
where we have included a factor $2$ for the reverse order of the
perturbations.
Evaluating this for the relevant diagram
\begin{equation}
V_4^{(b)}(2,2)={24J_1^2J_2^3 \over (4 {\tilde D})^4S}
\label{s7}
\end{equation}
where a factor $2$ for the mirror image process has been included.

Finally obtaining $V_3(2,2)$ from Eqs. (\ref{eqn:v4u}), (\ref{s1}),
(\ref{s3}), (\ref{s5}),and (\ref{s7}) and 
$V_3(2,3)$ from Eqs. (\ref{eqn:v4d}), (\ref{s2}), (\ref{s4}),and
(\ref{s6}) we are in a position to calculate the sum of the contributions to
$F(2,3)$ from all the diagrams of Fig. \ref{fig:v4}. We obtain
\begin{equation}
\delta F(2,3)=-{492 J_2^5 \over (4 {\tilde D})^4 S}
\end{equation}
where we have used $J_2=J_1/2+{\cal O}(1/ {\tilde D} S)$.
Combining this with Eqns. (\ref{eqn:v4u}) and (\ref{eqn:v4d}) we get
\begin{equation}
F(2,3)=-{392 J_2^5 \over (4 {\tilde D})^4 S} \ .
\end{equation}
This is negative showing that the $\langle 2 \rangle : \langle 3
\rangle$ phase boundary is indeed stable.

\section{Evaluation of ${\bf G}^n(i,j)$}

\subsection{Formulation for $G$}

The best way to get ${\bf G}^{(1,n)}(E)$, as defined
in Eq. (41), is as the inverse of the matrix
$Q_n \equiv [ E {\cal I} - S {\cal H}_0^{(1,n)} ]^{-1}$,
where ${\cal I}$ is the unit matrix.  We define

\begin{equation}
E+ 2 \tilde D +2J_1-2J_2 \equiv A \ , \ \ \ \ \ 
E + 2 \tilde D+2J_1-J_2 \equiv B \ \ \ \ \ E+ 2 \tilde D+J_1-J_2 = C \ .
\end{equation}
Then for $n=13$ the matrix ${\bf Q}_n$ is

\begin{equation} 
\label{QMAT}
{\bf Q}_{13} = \left[ \begin{array} { c c c c c c c c c c c c c}
C & -J_1 & J_2 & 0 & 0 & 0 & 0 & 0 & 0 & 0 & 0 & 0 & 0 \\
-J_1 &  B & -J_1 & J_2 & 0 & 0 & 0 & 0 & 0 & 0 & 0 & 0 & 0 \\
J_2  & -J_1 & A & -J_1 & J_2 & 0 & 0 & 0 & 0 & 0 & 0 & 0 & 0 \\
0  &  J_2 & -J_1 & A & -J_1 & J_2 & 0 & 0 & 0 & 0 & 0 & 0 & 0 \\
0  &  0 & J_2 & -J_1 & A & -J_1 & J_2 & 0 & 0 & 0 & 0 & 0 & 0 \\
0  &  0 & 0 & J_2 & -J_1 & A & -J_1 & J_2 & 0 & 0 & 0 & 0 & 0 \\
0  &  0 & 0 & 0 & J_2 & -J_1 & A & -J_1 & J_2 & 0 & 0 & 0 & 0 \\
0  &  0 & 0 & 0 & 0 & J_2 & -J_1 & A & -J_1 & J_2 & 0 & 0 & 0 \\
0  &  0 & 0 & 0 & 0 & 0 & J_2 & -J_1 & A & -J_1 & J_2 & 0 & 0 \\
0  &  0 & 0 & 0 & 0 & 0 & 0 & J_2 & -J_1 & A & -J_1 & J_2 & 0 \\
0  &  0 & 0 & 0 & 0 & 0 & 0 & 0 & J_2 & -J_1 & A & -J_1 & J_2 \\
0  &  0 & 0 & 0 & 0 & 0 & 0 & 0 & 0 & J_2 & -J_1 & B & -J_1 \\
0  &  0 & 0 & 0 & 0 & 0 & 0 & 0 & 0 & 0 & J_2 & -J_1 & C  \end{array}
\right]
\end{equation}
and

\begin{equation}
\label{GMAT}
G_{ij}^{(1,n)} (E) = \Biggl[ \Bigl( Q_{n} \Bigr)^{-1} \Biggr]_{ij}
\equiv (-1)^{i+j} {N}_n(i,j) / \Delta_n \ ,
\end{equation}
where $N_n(i,j)$ is the $(i,j)$ minor of ${\bf Q}_n$
and $\Delta_n = {\rm Det} {\bf Q}_n$.
Since  from inspection of Eq.\ (\ref{eqn:v2xwx}) $\Delta_n$ is thus a
common factor in $V_2(n)$, it is not 
necessary to carry out a complete evaluation of this quantity.
The ${\bf N}$'s are fortunately much easier to calculate.  The
first step, is to expand ${\bf N}$ by minors to get it in terms
of a matrix in which end effects, embodied in $B$ and $C$, have
been eliminated.  Thereby we obtain [writing $G$ for
$G^{(1,n)}(E)$]
\begin{mathletters}
\begin{eqnarray}
\label{GEQS}
(-1)^{n+1} \Delta_n G_{2,n-1} & = & C^2 d_{n-3} + 2CJ_1J_2 d_{n-4}
+ (2CJ_2^3+J_1^2J_2^2)d_{n-5} + 2J_1J_2^4 d_{n-6} + J_2^6 d_{n-7} \\
(-1)^{n+1} \Delta_n G_{1,n-1} & = & CJ_1 d_{n-3} + (J_1^2J_2 + BCJ_2) d_{n-4}
+ (J_1J_2^3 + BJ_1J_2^2 + CJ_1J_2^2) d_{n-5} \nonumber \\ && \
+ (CJ_2^4+BJ_2^4+J_1^2J_2^3) d_{n-6} +2J_1J_2^5 d_{n-7} + J_2^7 d_{n-8} \\
(-1)^{n+1} \Delta_n G_{1,n} & = &J_1^2 d_{n-3} + 2BJ_1 J_2 d_{n-4}
+ (2J_1^2J_2^2 + B^2J_2^2) d_{n-5} + (2BJ_1J_2^3+2J_1J_2^4) d_{n-6}
\nonumber \\ && +(2BJ_2^5+J_1^2 J_2^4) d_{n-7} + 2J_1J_2^6 d_{n-8}
+ J_2^8 d_{n-9} \ ,
\end{eqnarray}
\end{mathletters}
where $d_n$ denotes the determinant of the $n \times n$ matrix
${\bf D}_n$ which has no end effects:
\begin{equation} 
{\bf D}_{10} = \left[ \begin{array} { c c c c c c c c c c}
 -J_1 & A & -J_1 & J_2 & 0 & 0 & 0 & 0 & 0 & 0 \\
  J_2 & -J_1 & A & -J_1 & J_2 & 0 & 0 & 0 & 0 & 0 \\
  0 & J_2 & -J_1 & A & -J_1 & J_2 & 0 & 0 & 0 & 0 \\
  0 & 0 & J_2 & -J_1 & A & -J_1 & J_2 & 0 & 0 & 0 \\
  0 & 0 & 0 & J_2 & -J_1 & A & -J_1 & J_2 & 0 & 0 \\
  0 & 0 & 0 & 0 & J_2 & -J_1 & A & -J_1 & J_2 & 0 \\
  0 & 0 & 0 & 0 & 0 & J_2 & -J_1 & A & -J_1 & J_2 \\
  0 & 0 & 0 & 0 & 0 & 0 & J_2 & -J_1 & A & -J_1  \\
  0 & 0 & 0 & 0 & 0 & 0 & 0 & J_2 & -J_1 & A \\
  0 & 0 & 0 & 0 & 0 & 0 & 0 & 0 &J_2 & -J_1 \\
\end{array} \right] \ .
\end{equation}
From here on we will set $E=2 \tilde D + 2J_1 - 2J_2 \equiv 2D'$.

\subsection{Evaluation of  $d_n$}

Expanding the determinant of ${\bf D}$ about its upper left corner,
we have
\begin{equation}
\label{RECUR}
d_p = -J_1 d_{p-1} - J_2 A d_{p-2} - J_1J_2^2 d_{p-3} - J_2^4 d_{p-4} \ .
\end{equation}
We define
\begin{equation}
\label{RECURZ}
d_0 = 1
\end{equation}
and $d_p=0$ for $p<0$.  The solution of the recursion relation
is obtained from the result that
\begin{equation}
F(x) = \sum_{p=0}^\infty d_p x^p 
= \biggl( 1 + J_1x + J_2Ax^2 + J_1J_2^2 x^3 + J_2^4 x^4 \biggr)^{-1} 
\end{equation}
which, when expanded in powers of $x$ allows us to get $d_p$.
For this purpose we write
\begin{equation}
F(x) = \prod_{i=1}^4 \biggl( 1 - x J_2 y_i \biggr)^{-1} \ ,
\end{equation}
where $y_i$ is obtained as the solution to
\begin{equation}
\label{ROOT}
y_i^4 + (J_1/J_2)y_i^3 + (A/J_2)y_i^2 + (J_1/J_2) y_i + 1 =0 \ .
\end{equation}
We write the four roots of this equation as
\begin{equation}
y_1 =iy e^{i \delta} , \ \ \ \
y_2 =-iy e^{-i \delta} , \ \ \ \
y_3 = 1/y_1 = -iy^{-1} e^{-i \delta} \ , \ \ \ \ \
y_4 = 1/y_2 = iy^{-1} e^{i \delta} \ ,
\end{equation}
where $y$ and $\delta$ are positive, with
$y$ of order $\sqrt {A/J_2} \gg 1$ and $\delta \sim 1/y$.
Equation (\ref{ROOT}) can be solved as a quadratic equation for
$y_i/(y_i^2+1)$ from which $y_i$ can then be obtained.
Then $y$ and $\delta$ can be obtained as accurately as needed.

We now obtain $d_n$ by expanding $F(x)$ in powers of $x$.
To do that write
\begin{equation}
F(x)  = \biggl( \sum_{p=0}^\infty J_2^p x^p S_p \biggr)
\biggl( \sum_{r=0}^\infty J_2^r x^r T_r \biggr) \ ,
\end{equation}
where 
\begin{equation}
S_p = y_1^p + y_1^{p-1}y_2 \dots + y_2^p =
{y_1^{p+1} - y_2^{p+1} \over y_1 - y_2 } \equiv y^p Q_p \ .
\end{equation}
and
\begin{equation}
T_p = \Biggl[ {1 \over y_1^p} + {1 \over y_1^{p-1}} {1 \over y_2}
+ \dots {1 \over y_2^p } \Biggr] = {S_p \over y_1^p y_2^p } =
{S_p \over y^{2p} } = y^{-p} Q_p \ ,
\end{equation}
where

\begin{eqnarray}
\label{QEQ}
Q_p & =& i^p \biggl( e^{ip\delta} - e^{i(p-2)\delta } \dots
+ (-1)^p e^{-ip\delta } \biggr) \nonumber \\ & = &
(-1)^{p/2} { \cos [(p+1) \delta ] \over \cos \delta } \ ,
\ \ \ \ p \ \ {\rm even} \ ; \nonumber \\ &=&
(-1)^{(p+1)/2} { \sin [(p+1) \delta ] \over \cos \delta } \ ,
\ \ \ \ p \ \ {\rm odd} \ .
\end{eqnarray}
So
\begin{equation}
\label{DN}
d_n =  J_2^n y^n \sum_{r=0}^n Q_{n-r} Q_r y^{-2r} \ .
\end{equation}
To leading order in $J_2/A$, this gives
\begin{eqnarray}
d_p & = & J_2^p y^p (-1)^{p/2} \cos (p\delta) \ , \  \ \ \ \ 
\ \ \ \ p \ \ {\rm even} \ ; \nonumber \\ &=&
J_2^p y^p (-1)^{(p+1)/2}  \sin (p\delta ) \ , \ \ \ \ \
\ \ \ \ p \ \ {\rm odd} \ .
\end{eqnarray}

\subsection{Leading Evaluation of $V_2(n)$}

To leading order in $(J_2/D')$ Eq.\ (\ref{GEQS}) gives
\begin{equation}
G_{2,n-1} = (-1)^{n+1} {(4D')^2 \over \Delta_n } d_{n-3}\ .
\end{equation}
Asymptotically,
\begin{equation}
\Delta_n \sim J_2^n \lambda^n \sim \left( 4D' - {J_1^2 + J_2^2 \over
4D'} \dots \right)^n \equiv (4D')^n (1 - a)^n \ ,
\end{equation}
where $a$ is the correction of order $(J/D')^2$.
In fact, to leading order in $(J_2/D')$ we have
\begin{eqnarray}
G_{2,n-1} & = & {(4D')^2 \over \lambda^n}
(4 D' J_2)^{(n-3)/2} (-1)^{{n \over 2}}
\sin[(n-2)\delta ] \nonumber \\ &=&
{ \sqrt{4 D' } J_2^{-3/2} \over (1-a_0)^n }
\left( {J_2 \over 4D' } \right)^{n/2} (-1)^{(n/2)} \sin ( n \delta )
\ ,
\end{eqnarray}
for $n$ even, where $a_0 = 5(J_2/4 D')^2$ and
\begin{equation}
G_{2,n-1} = { \sqrt {4 D' } J_2^{-3/2} \over (1-a_0)^n }
\left( {J_2 \over 4D' } \right)^{n/2} (-1)^{(n-3)/2} \cos ( n \delta ) \ .
\end{equation}
for $n$ odd.  This result agrees with Eq.\ (\ref{V2RES}) to leading order
in $J_2/D'$.  For small $n$ these results reduce to those given
in Sec. III.

This evaluation is clearly not precise enough to tell whether
$V_2(n)$, Eq. (\ref{eqn:Vngreen}), is nonnegative.  Obviously, when $2n \delta /\pi$ is
close to an integer [or more precisely, when $G_{2,n-1}-2G_{1,n-1}=0$],
it is necessary to retain the first higher-order terms which are
nonzero there.  For this purpose we need to keep all the terms in
Eqs. (\ref{VAEQ}) and (\ref{VBEQ}).  Also in evaluating
the Green's functions we have to keep a sufficient number of
correction terms in Eq.\ (\ref{DN}).  The algebra required for this
analysis is too involved to be worth presenting.  Instead, we give
the final result in Eq. (\ref{V2RES}) and discuss how we have
numerically verified it.

\subsection{Numerical Verification of Eq. (54)}

Here we describe the comparison between the analytic results
of Eq.\ (\ref{V2RES}) and our numerical evaluation.  To clarify
the comparison we will work with rather large values of $D'/J_2$.
The numerical evaluation was done as follows.
In all cases we set $J_1=2J_2$.  First we obtained $y$ and $\delta$
exactly by solving Eq. (\ref{ROOT}).  Then we constructed the $d_n$
using Eq. (\ref{DN}).  We checked that the $d_n$ so obtained did
satisfy the recursion relation of Eq. (\ref{RECUR}) to one part
in $10^{10}$.  Next we constructed the quantities $\Delta_n G_{ij}$
using Eqs. (B4).  To calculate $V_2(n)$ according to Eq. (\ref{V2EQ})
we set $\Delta_n=J_2^n \lambda^n$ and used the
approximation $\lambda \approx \lambda_0$, where
\begin{equation}
\lambda_0 = {4D' \over J_2 } \Biggl( 1 - {5J_2^2 \over 16 D'^2} \Biggr) \ .
\end{equation}
This approximation only affects slightly the scale of $V_2(n)$ because
all the $G$'s appearing in Eq. (\ref{V2EQ}) are proportional to
this factor.

To numerically verify Eq. (\ref{V2RES}) we chose to study
$n$ odd, because this case is the first where $V_2(n)$ approaches
zero.  In particular we will explicitly discuss only the case
of $n=63$ and $D'/J_2=400$ for which
$n \sqrt {J_2 /(4\tilde D)} \approx \pi /2$.
For the parameters we used, $V_2(n)$ became very small.
This is because to check the asymptotic forms it is convenient
to assign values to the parameters well outside anything one
would encounter experimentally.  For instance,
$V_2(n)/J_2$ became of order $10^{-200}$.  Obviously, to interpret
the numerical results, it is convenient to consider quantities
in which the exponential decay [$\lambda^{-n}$ in Eq. (\ref{V2RES})]
is removed.  Accordingly, we list in Table 1 the
values of 
\begin{equation}
\label{F1N}
F_1^{\rm exact} \equiv J_2^{2-2n} \Delta_n^2 \lambda_0^{-n}
\Bigl( G_{2,n-1} - 2 G_{1,n-1} \Bigr)^2 \ ,
\end{equation}
\noindent for a few representative cases where $F_1^{\rm exact}$ is close to
zero.
The subscript 'exact' indicates that we evaluated this quantity to
double-precision accuracy.  (Recall that our precise evaluation is
for $\Delta_n G$ and not for $G$ itself.) We can write $F_1$ as
\begin{equation}
F_1^{\rm exact} = {1 \over 4} S \Delta_n^2 \lambda_0^{-n} 
J_2^{-2n-1} V_A = {1 \over 4} (\lambda
/ \lambda_0 )^n (S V_A \lambda^n /J) \ .
\end{equation}
Since $\lambda \approx \lambda_0$ and $V_a \lambda^n$ is of order
unity, $F_1^{\rm exact}$ will be of order unity as desired.
According to Eq. (\ref{V2RES}) whose validity we wish to verify,
we have the analytic result for $F_1$:
\begin{equation}
\label{F1}
F_1^{\rm analytic} = \left( {\lambda \over \lambda_0 } \right)^n
{4 D' \over J_2 }
\cos^2 ( n \delta + 4 \delta^3 ) \approx {4D' \over J_2}
\cos^2 ( n \delta + 4 \delta^3 ) \ .
\end{equation}
This analytic result predicts that $F_1$ should become zero
when $\delta= \delta_0$, where $n \delta_0 + 4 \delta_0^3 = \pi /2$,
from which we get $\delta_0 = 0.0249322909$.  From the solution to
Eq. (\ref{ROOT}) we have that
\begin{equation}
\sqrt {J_2/A} = \delta + (5/6) \delta^3 \ ,
\end{equation}
which gives $A/J_2= 1607.037$ or, since $A=4 \tilde D + 4 J_2$,
$\tilde D / J_2 = 400.759$, in very precise agreement with the
the location of the zero of $V_A$ from the numerical evaluation
given in Table 1.  So we have confirmed that that $F_1$ is proportional to
$\cos ( n\delta + 4 \delta^3 )$.

Now we turn to the evaluation of $V_B(n)$.  We are especially
interested in $V_B$ at the point when $V_A(n)=0$.
To study this quantity we have listed in Table 1 values of
\begin{eqnarray}
\label{F2}
& &F_2^{\rm exact} \equiv J_2^{2-2n} \Delta_n^2 \lambda_0^{-n}
\Bigl( 4G_{1,n}G_{1,n-1} - 5 G_{1,n}^2 \Bigr) \nonumber \\
& & = {1 \over 4} \left( { \lambda \over \lambda_0 } \right)^n
(SV_B \lambda^n /J )  \ .
\end{eqnarray}
A numerical evaluation to double precision accuracy at the point where
$V_A=0$ is given in Table 1 as
\begin{equation}
F_2^{\rm exact} = -1.004 \times 10^{-5} \ .
\end{equation}
The analog of Eq. (\ref{F1}) is
\begin{equation}
\label{F2AN}
F_2^{\rm analytic} = -(3/2) (\lambda / \lambda_0)^n (J_2 / \tilde D)^2 \ .
\end{equation}
In contrast to the previous check, here we actually need to rely
on the approximations for the various scale factors.  Assuming
$\lambda \approx \lambda_0$, the above equation the above equation
gives $F_2^{\rm analytic} = -0.9 \times 10^{-5}$, compared
to the exact result.  This comparison may not seem impressive.
However, it does indicate that we have assigned the correct order
in $J_2/\tilde{D}$ to $V_B$.  One notices that $F_2$ does depend on $\tilde D$.
An empirical fit to the numerical data indicates that a better
approximation for $V_B$ (when $V_A \approx 0$) is to write
\begin{equation}
V_B = - {6\tilde D \over S \lambda_n} \Biggl[
\left( {J_2 \over \tilde D } \right)^3
+ 14 \sin (2n \delta + 8 \delta^3 ) \left( {J_2 \over \tilde D }
\right)^2 \Biggr] \ .
\end{equation}
Therefore we replaced Eq. (\ref{F2AN}) by
\begin{equation}
F_2^{\rm analytic} = - 24 \left( {J_2 \over 4 \tilde D } \right)^2
\left( { \Delta_n^2 \over J_2^{2n} \lambda^n \lambda_0^n } \right)
\Biggl( 1 + 14 (\tilde D / J_2) \sin (2n \delta + 8 \delta^3 ) \Biggr) \ .
\label{NEW1}
\end{equation}
Note that Eqs. (\ref{F2AN}) and (\ref{NEW1}) give the same result when
$F_1=0$ because then $\sin (2n\delta+8\delta^3)=0$. However, Eq. (\ref{F2AN})
reproduces the variation in $F_2$ when $F_1$ is not exactly zero.

To see an overall comparison between
numerical and analytic results we also tabulate the ratios
\begin{equation}
\label{RR}
R_i = F_i^{\rm exact} / F_i^{\rm analytic} \ ,
\end{equation}
\noindent where we used Eq. (B31) for $F_2^{\rm analytic}$.
It is striking that although the quantities $F_1^{\rm exact}$ and $F_2^{\rm exact}$
vary over many decades, that the ratios $R_i$ between
numerical and analytic results are essentially constant, except
very near where the quantities pass through zero and a small
error in the phase shift can cause large variations in these
ratios.  From this discussion we conclude that our numerical
results corroborate the complicated algebra leading to
Eq. (\ref{V2RES}).

\begin{table}
\caption{Numerical evaluations of $V_2$.  Here $F=F_1+F_2$, where
$F_1$ and $F_2$ are defined in Eqs. (\protect\ref{F1N}) and
(\protect\ref{F2}). Also $R_i$ is defined in Eq. (\protect\ref{RR}).
The data for $n=95$ is for $n \delta \approx 3 \pi /2$.  All the other
data is for $n \delta \approx \pi /2$.}
\begin{tabular} {  c  r r r r r r}
$\tilde D/J_2$ & $n$ & $ F \ \ \ \ \ \ \ \ $ & $F_1^{\rm exact} \ \ \ \
$ & $F_2^{\rm exact}$\ \ \ \ & $R_1 \ \ $ & $R_2 \ \ $ \\
    99.000&  95&     .5223948307&     .5134465900&     .0089482407&   1.611&    .759\\  
   100.000&  95&     .0165890715&     .0152594233&     .0013296482&   1.603&    .705\\  
   100.100&  95&     .0047456479&     .0041642309&     .0005814170&   1.602&    .637\\  
   100.180&  95&     .0002864392&     .0003018071&    -.0000153679&   1.599&   -.112\\  
   100.190&  95&     .0000419448&     .0001317981&    -.0000898533&   1.598&  -2.200\\  
   100.200&  95&    -.0001330808&     .0000312329&    -.0001643137&   1.593&   2.939\\  
   100.205&  95&    -.0001945491&     .0000069855&    -.0002015345&   1.583&   1.933\\  
   100.206&  95&    -.0002047595&     .0000042185&    -.0002089779&   1.578&   1.834\\  
   100.207&  95&    -.0002142755&     .0000021456&    -.0002164211&   1.568&   1.751\\  
   100.208&  95&    -.0002230972&     .0000007668&    -.0002238640&   1.546&   1.680\\  
   100.209&  95&    -.0002312245&     .0000000821&    -.0002313067&   1.440&   1.618\\  
   100.210&  95&    -.0002386576&     .0000000915&    -.0002387491&   1.782&   1.565\\  
   100.211&  95&    -.0002453963&     .0000007949&    -.0002461912&   1.660&   1.517\\  
   100.212&  95&    -.0002514408&     .0000021923&    -.0002536331&   1.636&   1.475\\  
   100.213&  95&    -.0002567910&     .0000042838&    -.0002610748&   1.626&   1.438\\  
   100.214&  95&    -.0002614470&     .0000070692&    -.0002685162&   1.621&   1.404\\  
   100.215&  95&    -.0002654088&     .0000105486&    -.0002759573&   1.617&   1.373\\  
   100.220&  95&    -.0002748052&     .0000383541&    -.0003131593&   1.610&   1.256\\  
   100.300&  95&     .0019330782&     .0028406168&    -.0009075386&   1.602&    .889\\  
   100.500&  95&     .0268205406&     .0292069915&    -.0023864509&   1.600&    .811\\  
   101.000&  95&     .2091773178&     .2152168605&    -.0060395426&   1.596&    .786\\ \hline 
   765.000&  87&     .0009103506&     .0008733126&     .0000370379&   1.058&   1.429\\  
   765.499&  87&    -.0000018986&     .0000002010&    -.0000020996&   1.057&    .988\\  
   765.500&  87&    -.0000020258&     .0000001521&    -.0000021779&   1.057&    .999\\  
   765.501&  87&    -.0000021463&     .0000001100&    -.0000022563&   1.057&   1.009\\  
   765.502&  87&    -.0000022599&     .0000000747&    -.0000023346&   1.056&   1.018\\  
   766.000&  87&     .0007858863&     .0008272135&    -.0000413271&   1.058&   1.367\\  \hline 
   508.000&  71&     .0099390222&     .0097440109&     .0001950113&   1.072&   1.155\\  
   509.000&  71&     .0007640012&     .0007157580&     .0000482432&   1.072&   1.165\\  
   509.345&  71&     .0000015730&     .0000037895&    -.0000022165&   1.071&    .947\\  
   509.350&  71&    -.0000004279&     .0000025192&    -.0000029471&   1.071&    .991\\  
   509.378&  71&    -.0000068571&     .0000001812&    -.0000070383&   1.074&   1.080\\  
   509.379&  71&    -.0000069369&     .0000002476&    -.0000071844&   1.074&   1.081\\  
   509.380&  71&    -.0000070063&     .0000003243&    -.0000073305&   1.074&   1.082\\  
   509.381&  71&    -.0000070653&     .0000004113&    -.0000074766&   1.073&   1.084\\  
   509.400&  71&    -.0000062236&     .0000040289&    -.0000102525&   1.072&   1.102\\  
   509.450&  71&     .0000138198&     .0000313758&    -.0000175559&   1.072&   1.123\\  
   509.600&  71&     .0002289421&     .0002683972&    -.0000394551&   1.072&   1.140\\  
   510.000&  71&     .0019384519&     .0020362219&    -.0000977700&   1.072&   1.149\\  \hline 
   390.000&  63&     .7804778656&     .7781343381&     .0023435276&   1.083&   1.021\\  
   395.000&  63&     .2225976723&     .2213686283&     .0012290440&   1.082&   1.026\\  
   400.000&  63&     .0039724306&     .0038217672&     .0001506633&   1.081&   1.029\\  
   400.500&  63&     .0004905488&     .0004457817&     .0000447672&   1.081&   1.023\\  
   400.600&  63&     .0001919976&     .0001683678&     .0000236298&   1.081&   1.014\\  
   400.700&  63&     .0000259217&     .0000234154&     .0000025063&   1.081&    .881\\  
   400.710&  63&     .0000165981&     .0000162033&     .0000003948&   1.081&    .493\\  
   400.720&  63&     .0000085986&     .0000103153&    -.0000017167&   1.081&   1.379\\  
   400.730&  63&     .0000019232&     .0000057512&    -.0000038280&   1.080&   1.164\\  
   400.740&  63&    -.0000034282&     .0000025109&    -.0000059391&   1.080&   1.114\\  
   400.750&  63&    -.0000074557&     .0000005945&    -.0000080501&   1.079&   1.091\\  
   400.757&  63&    -.0000094872&     .0000000406&    -.0000095278&   1.072&   1.082\\  
   400.758&  63&    -.0000097244&     .0000000144&    -.0000097389&   1.067&   1.081\\  
   400.759&  63&    -.0000099484&     .0000000015&    -.0000099499&   1.037&   1.080\\  
   400.760&  63&    -.0000101592&     .0000000018&    -.0000101610&   1.123&   1.079\\  
   400.761&  63&    -.0000103568&     .0000000153&    -.0000103721&   1.095&   1.078\\  
   400.762&  63&    -.0000105411&     .0000000421&    -.0000105832&   1.090&   1.077\\  
   400.770&  63&    -.0000115389&     .0000007328&    -.0000122718&   1.083&   1.070\\  
   400.780&  63&    -.0000115948&     .0000027875&    -.0000143824&   1.082&   1.065\\  
   400.790&  63&    -.0000103270&     .0000061658&    -.0000164928&   1.082&   1.061\\  
   400.800&  63&    -.0000077355&     .0000108676&    -.0000186032&   1.082&   1.057\\  
   400.900&  63&     .0000909688&     .0001306676&    -.0000396988&   1.081&   1.044\\  
   401.000&  63&     .0003219780&     .0003827584&    -.0000607804&   1.081&   1.040\\  
   402.000&  63&     .0098963366&     .0101671709&    -.0002708343&   1.081&   1.036\\  
   403.000&  63&     .0326387797&     .0331182868&    -.0004795072&   1.081&   1.036\\  
   404.000&  63&     .0684928015&     .0691796094&    -.0006868078&   1.080&   1.037\\  
   405.000&  63&     .1174021181&     .1182948632&    -.0008927450&   1.080&   1.038\\  
   410.000&  63&     .5558237738&     .5577260578&    -.0019022840&   1.079&   1.043\\  
\end{tabular}
\end{table}

\section{Analysis of Eq. (\ref{V2GEN}) when One Term Dominates}

Here we analyze Eq. (\ref{V2GEN}) in the limit when $n$ is so large that
only a small range of $\alpha$ is important.  Superficially,
if only a single value of $\alpha$ were important, one could make
$V_2(n)$ negative by adjusting $J_2/D$ so that the first square
bracket vanished.  We now show that this reasoning is incorrect.
The argument is most easily described when one arbitrarily sets
$\phi_\alpha(0)=1$ and $\phi_\alpha(-1)=0$.  Since these quantities
depend only weakly on $\alpha$, this simplification is only a matter
of convenience.  The crucial $\alpha$-dependence
is that in $G_n( \epsilon_\alpha^{(-\infty,0)})$ when $n$ is
arbitrarily large.  For that regime we treat the case when the
contribution to $V_2(n)$ comes from near where the summand is maximal.  
Therefore
we have [writing $\epsilon_\alpha$ for $\epsilon_\alpha^{(-\infty,0)}$]
\begin{equation}
\epsilon_\alpha = \epsilon_{\alpha_0} + 4 \gamma (\alpha-\alpha_0)^2
\equiv 2D'' + 4 \gamma x^2 \ ,
\end{equation}
where $D''$ is of order $D$ and $x=\alpha-\alpha_0$.  Thus, when the
sum over $\alpha$ is replaced by an integral, Eq. (\ref{V2GEN}) (for $n$ even) becomes
\begin{equation}
\label{C2}
V_2(n) \sim \int dx \left( {J_2 \over 4D + 4\gamma x^2 } \right)^n
\Biggl[ \sin^2 [ n \sqrt{ J_2 / (4D+ 4\gamma x^2 )} ]  -  C (J_2/D)^3
\Biggr] \ ,
\end{equation}
where we used the result of Eq. (\ref{V2RES}) to write the negative
correction term.  Also from Appendix B we
identified $4 D'$ as begin $\epsilon_\alpha +2D'$ and we dropped
the distinction between $D'$ and $D''$. For large $n$ we write
\begin{eqnarray}
V_2(n) & \sim & \left( {J_2 \over 4D} \right)^n \int_{-\infty}^\infty
dx e^ {- n \epsilon x^2 } {\rm Re} \Biggl\{ 1 -
e^{in \sqrt {J_2/(D + \gamma x^2 )}}  -  2C (J_2/D)^3 \Biggr\}
\nonumber \\ &=& \left( {J_2 \over 4D} \right)^n \Biggl[ \Biggl(
1 - 2C (J_2/D)^3 \Biggr) \Biggl( { \pi \over n \epsilon } \Biggr)^{1/2} 
- {\rm Re} \Biggl\{ e^{in \sqrt {J_2/D} } \int_{-\infty}^\infty
dx e^{-n \tilde \epsilon x^2 + \eta x^4 \dots } \Biggr\} \Biggr] \ ,
\end{eqnarray}
where $\epsilon=\gamma/D$, $\tilde \epsilon = \epsilon - (i \gamma/2)
\sqrt {J_2/D^3}$, and $\eta = (3/4)in\gamma^2 \sqrt{ J_2/D^5}$.  For
large $n$ we can drop the term in $\eta$, so that
\begin{eqnarray}
V_2 (n) & = & \left( {J_2 \over 4D} \right)^n \sqrt {\pi /(n\epsilon)}
\Biggl[ 1 - 2 C (J_2/D)^3 - {\rm Re} \Biggl\{  e^{in \sqrt {J_2/D} }
\sqrt { \epsilon / \tilde \epsilon } \Biggr\} \Biggr] \nonumber \\
&=& \left( {J_2 \over 4D} \right)^n \sqrt {\pi /(n\epsilon)}
\Biggl[ 1 - 2 C (J_2/D)^3 - {\rm Re} \Biggl\{  e^{in \sqrt {J_2/D}}
{1 \over \sqrt { 1 - {i \over 2} \sqrt{J_2 \over D } } }
\Biggr\} \Biggr] \ . 
\end{eqnarray}
Thus
\begin{eqnarray}
V_2(n) \left( {4D \over J_2 } \right)^n   \sqrt { {n \epsilon \over \pi }}
&>& 1 - 2C(J_2/D)^3 - {1 \over \mid 1 - {i \over 2} \sqrt{ {J_2 \over D}}
\mid^{1/2}  } \nonumber \\ &=&
1 - 2 C(J_2/D)^3 - {1 \over \left( 1 + {J_2 \over 4D } \right)^{1/4} }
\approx  {J_2 \over 16D } - C \left( {J_2 \over D } \right)^3 \ .
\end{eqnarray}
Thus $V_2(n)$ is positive in the limit of asymptotically large $n$.
It is easy to see from Eq. (\ref{C2}) that for large $n$ the variable
$x$ can be of order $\sqrt{D/(n\gamma)}$, which can cause a variation in
the argument of the sine function of order $\sqrt {J_2/D}$.
This estimate immediately explains the final result.

\vspace{1.0cm}

\vspace{1.0cm}

\vspace{1.0cm}

\vspace{1.0cm}

\vspace{1.0cm}

\vspace{1.0cm}

\vspace{1.0cm}

\vspace{1.0cm}

\vspace{1.0cm}

\vspace{1.0cm}

\vspace{1.0cm}

\vspace{1.0cm}

\vspace{1.0cm}

\vspace{1.0cm}
\vspace{1.0cm}

\vspace{1.0cm}

\vspace{1.0cm}


\begin{references}

\bibitem{JJ}
{\it Rare Earth Magnetism Structures and Excitations}. J. Jensen and
A. R. Mackintosh (Oxford University Press, Oxford 1991).
\bibitem{alloys}
A. Loiseau, G Van Tendeloo, R Portier and F Ducastelle,
J. Phys. (Paris) {\bf 46}, 595 (1985).

\bibitem{politypism}
P. Krishna (ed.), J. Cryst. Growth Charact. {\bf 7} (1984).


\bibitem{MEFXS}
M. E. Fisher and A. M. Szpilka, Phys. Rev. B {\bf 36}, 5343 (1987);\ 
A. M. Szpilka and M. E. Fisher,
Phys. Rev. Lett. {\bf 57}, 1044 (1986).

\bibitem{RJE}
R. J. Elliott, Phys. Rev. {\bf 124}, 346 (1961); J. M. Yeomans, in {\it Solid State Physics},
{\bf 41} (H. Ehrenreich and D. Turnbull eds, Academic Press, Orlando, 1988); W. Selke, Phys. Report {\bf
170}, 213 (1988); W. Selke in {\it Phase transitions and Critical
Phenomena} vol. 15, eds C. Domb and J. L. Lebowitz, (New York:
Academic, 1992).

\bibitem{BAK}
P. Bak and J. von Boehm, Phys. Rev. B.{\bf 21}, 5297 (1980).

\bibitem{MEFWS}
M. E. Fisher and W. Selke, Phys. Rev. Lett. {\bf 44}, 1502 (1980);
M. E. Fisher and W. Selke, Philos. Trans. R. Soc. (London), {\bf A
302}, 1 (1981).

\bibitem{VG}
J. Villain and M. B. Gordon, J. Phys. C. {\bf 13}, 3117 (1980).

\bibitem{EFS}
E. F. Shender, Soviet Physics, JETP {\bf 56}, 178 (1982).

\bibitem{CLH}
C. Henley, Phys. Rev. Lett. {\bf 62}, 2056 (1989).

\bibitem{GO}
M. J. de Oliveira and R. B. Griffiths, Surf. Sci. {\bf 71}, 687 (1978).

\bibitem{unp} C. Micheletti, unpublished.

\bibitem{DM}
F. J. Dyson, Phys. Rev. {\bf 102}, 1217, 1230 (1956).

\bibitem{SVM}
S. V. Maleev, Sov. Phys.  -- JETP {\bf 6}, 776 (1958).

\bibitem{NEGLECT}
As Dyson has shown [\onlinecite{DM}] the effect of spin-wave interactions
($V_4$) is essentially proportional to the density of spin excitations.
In fact, $V_4$ has NO effect on states containing a single spin 
excitation.  On these grounds we expect that spin-wave interactions
can be neglected in the present context.

\bibitem{DuxY}
P. M. Duxbury and J. M. Yeomans, J. Phys {\bf A 18}, L983 (1985).

\bibitem{Messiah} 
{\it Quantum Mechanics} A. Messiah. (North-Holland,
Amsterdam 1966 ).

\bibitem{HMY}
A. B. Harris, C. Micheletti and J. M. Yeomans, {\em Quantum
Fluctuations in the ANNNI Model}, to be published in Physical Review Letters.


\bibitem{RBG}
K. E. Bassler, K. Sasaki, and R. B. Griffiths, J. Stat. Phys.
{\bf 62}, 45 (1991).

\bibitem{SY} F. Seno and J. M. Yeomans, Phys. Rev. B 50, 10385 (1994)

\bibitem{MY} C. Micheletti and J. M. Yeomans, Europhys. Lett. {\bf
28}, 465 (1994).

\bibitem{RMP} R. J. Elliott, J. A. Krumhansl, and P. L. Leath,
Rev. Mod. Phys. {\bf 46}, 465 (1974).

\end{references}
\end{document}